\documentclass[11pt]{article}
\pdfoutput=1
\usepackage[centertags]{amsmath}
\usepackage[square, comma, sort&compress,numbers]{natbib}
\usepackage{array,multirow}
\numberwithin{equation}{section}
\usepackage{amssymb}
\usepackage{graphicx}
\usepackage{color}
\usepackage[normalem]{ulem}

\usepackage{epsfig}
\usepackage{bbold}

\usepackage{wrapfig}
\usepackage{float}
\usepackage{soul}

\usepackage{tikz,pgf}
\usetikzlibrary{shapes}
\usetikzlibrary{calc}
\usetikzlibrary{decorations.pathmorphing}
\usetikzlibrary{decorations.pathreplacing,shapes.misc}
\usetikzlibrary{positioning}
\usetikzlibrary{arrows}
\usetikzlibrary{decorations.markings}
\usetikzlibrary{shadings}

\usetikzlibrary{intersections}
\usepackage{mathrsfs}
\usepackage{mathtools}
\usepackage{cancel}
\usepackage{tikz-cd}

\usepackage{tikz}
\usepackage{cases}
\def\sideremark#1{\ifvmode\leavevmode\fi\vadjust{\vbox to0pt{\vss
 \hbox to 0pt{\hskip\hsize\hskip1em
 \vbox{\hsize3cm\tiny\raggedright\pretolerance10000
  \noindent #1\hfill}\hss}\vbox to8pt{\vfil}\vss}}}

\def\be{\begin{equation}}
\def\ee{\end{equation}}
\def\ba{\begin{array}}
\def\ea{\end{array}}

\def\dps{\displaystyle}

\advance\textheight\topskip
\renewcommand{\tilde}{\widetilde}
\renewcommand{\hat}{\widehat}

\newcommand{\bref}[1]{\textbf{\ref{#1}}}

\newcommand{\dd}{\partial}

\renewcommand{\geq}{\,{\geqslant}\,}
\renewcommand{\leq}{\,{\leqslant}\,}

\newcommand{\binner}[2]{%
  {\langle}\kern-4.15pt{\langle}#1{,}\,#2{\rangle}\kern-4.15pt{\rangle}}

\newcommand{\half}{\mathchoice{%
    \ffrac{1}{2}}{\frac{1}{2}}{\frac{1}{2}}{\frac{1}{2}}}

\newcommand{\ffrac}[2]{\raisebox{.5pt}%
  {\footnotesize$\displaystyle\frac{#1}{#2}$}\kern1pt}

\def\cD{\mathcal{D}}

\def\cG{\mathcal{G}}
\def\cH{\mathcal{H}}

\def\cM{\mathcal{M}}

\def\cS{\mathcal{S}}

\def\cV{\mathcal{V}}

\numberwithin{equation}{section} \makeatletter
\@addtoreset{equation}{section}

\def\be{\begin{equation}}
\def\ee{\end{equation}}
\def\ba{\begin{array}}
\def\ea{\end{array}}

\def\dps{\displaystyle}

\def\1{\tilde{1}}
\def\2{\tilde{2}}
\def\3{\tilde{3}}

\newdimen\tableauside\tableauside=1.0ex
\newdimen\tableaurule\tableaurule=0.4pt
\newdimen\tableaustep
\def\phantomhrule#1{\hbox{\vbox to0pt{\hrule height\tableaurule
width#1\vss}}}
\def\phantomvrule#1{\vbox{\hbox to0pt{\vrule width\tableaurule
height#1\hss}}}
\def\sqr{\vbox{%
  \phantomhrule\tableaustep

\hbox{\phantomvrule\tableaustep\kern\tableaustep\phantomvrule\tableaustep}%
  \hbox{\vbox{\phantomhrule\tableauside}\kern-\tableaurule}}}
\def\squares#1{\hbox{\count0=#1\noindent\loop\sqr
  \advance\count0 by-1 \ifnum\count0>0\repeat}}
\def\tableau#1{\vcenter{\offinterlineskip
  \tableaustep=\tableauside\advance\tableaustep by-\tableaurule
  \kern\normallineskip\hbox
    {\kern\normallineskip\vbox
      {\gettableau#1 0 }%
     \kern\normallineskip\kern\tableaurule}%
  \kern\normallineskip\kern\tableaurule}}
\def\gettableau#1 {\ifnum#1=0\let\next=\null\else
  \squares{#1}\let\next=\gettableau\fi\next}

\def\cD{\mathcal{D}}

\def\cG{\mathcal{G}}
\def\cH{\mathcal{H}}

\def\cM{\mathcal{M}}

\def\cS{\mathcal{S}}

\def\cV{\mathcal{V}}

\numberwithin{equation}{section} \makeatletter
\@addtoreset{equation}{section}

\def\be{\begin{equation}}
\def\ee{\end{equation}}
\def\ba{\begin{array}}
\def\ea{\end{array}}

\def\dps{\displaystyle}

\def\ba{\begin{array}}
\def\ea{\end{array}}

\def\dps{\displaystyle}

\def\3ads{AdS$_3$}

\def\2ads{AdS$_2$}

\def\hdegr{AdS$_{d+1}$/AdS$_d$ }

\def\ldegr{AdS$_{3}$/AdS$_2$ }

\def\lao22{\text{o}(2,2)}
\def\cao2{\text{o}(2)}
\def\ao12{\text{o}(1,2)}

\def\lgo22{\text{O}(2,2)}
\def\lgop22{\text{O}^+(2,2)}

\def\algo22{\text{O}(2,1)}
\def\algop22{\text{O}^+(2,1)}

\def\nD{\cH}

\newenvironment{al}
    {\begin{equation}
    \begin{aligned}
    }
    {
    \end{aligned}%
    \end{equation}%
    }

\newcommand{\ol}[1]{\overline{#1}}
\newcommand{\mm}[2][2]{\gamma_{#1|#2}}
\newcommand{\LL}{\L}

\DeclareMathOperator{\sech}{sech}

\usepackage[tracking=true]{microtype}

\newcommand\narrowstyle{\SetTracking{encoding=*}{-50}\lsstyle}
\newcommand\normalstyle{\SetTracking{encoding=*}{0}\lsstyle}

\DeclarePairedDelimiter\abs{\lvert}{\rvert}%

\usepackage{jheppub}

\makeatletter
\def\@fpheader{\vspace{-.1cm}}
\makeatother

\title{AdS$_3$/AdS$_2$  degression of massless particles}

\author[a,b]{Konstantin\ Alkalaev} 
\author[b]{and Alexander\ Yan}

\affiliation[a]{I.E. Tamm Department of Theoretical Physics, \\P.N. Lebedev Physical
Institute, 119991 Moscow, Russia}
\affiliation[b]{Institute for Theoretical and Mathematical Physics,\\
Lomonosov Moscow State University,
119991 Moscow, Russia}

\emailAdd{alkalaev@lpi.ru, yan@lpi.ru}

\abstract{We study a 3d/2d dimensional degression which is a Kaluza-Klein type mechanism in \3ads space foliated into \2ads hypersurfaces. It is shown that  an AdS$_3$ massless particle of spin $s=1,2,..., \infty$ degresses into a couple of  AdS$_2$ particles of equal energies $E=s$. Note that the Kaluza-Klein spectra in higher dimensions   are always infinite. To formulate the  \ldegr degression we consider  branching rules for \3ads isometry algebra \lao22 representations decomposed with respect to \2ads isometry algebra \ao12. We find  that a given   \lao22  higher-spin representation lying on the unitary bound (i.e. massless) decomposes into two equal  \ao12 modules. In the field-theoretical terms, this phenomenon is demonstrated  for spin-2 and spin-3 free massless fields. The truncation to a finite spectrum can be seen by using particular mode expansions, (partial) diagonalizations, and identities specific to two dimensions.        
}

\begin{document}

\maketitle
\flushbottom

\allowdisplaybreaks
%\allowdisplaybreaks[0]
\section{Introduction}

A distinctive feature of Kaluza-Klein compactification \cite{Kaluza:1921tu,Klein:1926tv} of a  top theory  are  infinite towers of fields of increasing masses and decreasing spins in the resulting bottom theory.\footnote{In this paper, we use the terms {\it top} and {\it bottom} theories to designate respective  higher-dimensional and lower-dimensional theories in the context of  the Kaluza-Klein type mechanism.} The same property holds both in the AdS dimensional  degression \cite{Metsaev:2000qb,Artsukevich:2008vy} and the AdS waveguide compactification \cite{Gwak:2016sma}. In this paper we show that the spectrum of a bottom theory can be finite if one starts with a {\it topological} theory in higher-dimensional AdS space. Specifically, we study the \ldegr degression for a spin-$s$ massless field theory. The resulting \2ads spectrum consists of finitely many {\it propagating} modes. To some extent this phenomenon is similar to the Brown-Henneaux relation for $3d$ Einstein  gravity with the cosmological term without local degrees of freedom which under appropriate AdS$_3$ boundary conditions 
\narrowstyle 
leads to $2d$ boundary $\mathbb{R}^2$ local degrees of freedom spanning Virasoro algebra representations \cite{Brown:1986nw}.     
\normalstyle

In the \hdegr degression the spectrum of a bottom AdS$_{d}$ theory is defined by branching rules for a particular representation of the AdS$_{d+1}$ isometry algebra  $\text{o}(d,2)$  with respect to the AdS$_d$ isometry algebra $\text{o}(d-1,2)\subset \text{o}(d,2)$. The $\text{o}(d,2)$ representation may be chosen to describe   massive or (partially-)massless AdS$_{d+1}$ particle of any spin. Then, in $d>2$  the branching rules for a massless spin-$s$ $\text{o}(d,2)$ representation  give the following general pattern \cite{Artsukevich:2008vy}: an infinite upper spin-$s$ sequence of $\text{o}(d-1,2)$ representations with running energies and a finite collection of lower spin-$q$  $\text{o}(d-1,2)$ representations with fixed energies, where $q=0,1,...,s-1$. In the \ldegr degression the branching rules for relevant  $\text{o}(2,2)$ representations with respect to $\text{o}(1,2)$ subalgebra are drastically simplified,  involving  only finitely many $\text{o}(1,2)$ representations. Namely, a massless  $\text{o}(2,2)$ representation of spin $s = 1,2,3,...$ turns out to be  isomorphic to two $\text{o}(1,2)$ representations which are Verma modules of equal weights $s$.     

On the field-theoretical level, when considering respective field theory of higher rank tensor (gauge) fields,  such a truncation is basically due to the so-called Schouten identities for two-dimensional kinetic operators. It follows that a massless spin-$s$ \3ads top theory degresses into an \2ads bottom theory  consisting of the Klein-Gordon and  Proca fields of equal energies $E=s$ (but different masses $m^2_{KG}=E(E-1)$ and $m^2_P=E(E-1)-1$, respectively). On the contrary, degressing spin-0 massive/massless  theories  in \3ads which do have local degrees of freedom we still obtain an infinite spectrum of the respective bottom theory in  \2ads.\footnote{Discussion of AdS$_3$ field dynamics  and review of the SL$(2, \mathbb{R})$ group representations  can be found e.g. in  \cite{Balasubramanian:1998sn,Metsaev:2000qb}. AdS$_2$ field equations and their solutions attracted recently some attention mainly because of the SYK/JT duality problem and the conformal bootstrap, see e.g. recent \cite{Mazac:2016qev,Anninos:2019oka,Anous:2020nxu}.  
} In general, an infinite \2ads spectrum persists for all massive spin-$s$ field \3ads theories.

The paper is organized as follows. In Section \bref{sec:brief} we discuss the general known facts about compactification/degression in AdS spaces. We summarize the basic features of the \ldegr degression focusing on differences with the higher-dimensional case which result in truncating the spectrum from infinite to finite one.  In Section \bref{sec:branch} we shortly recall a few basic facts about relevant $\text{o}(d,2)$ representations paying particular attention to $d=1$ and $d=2$ cases.\footnote{For more discussion in the present context see e.g. \cite{Boulanger:2014vya,Gwak:2015jdo,Basile:2018dzi,Alkalaev:2019xuv} and Appendix \bref{app:dec}.} Also, we formulate the branching rules specifying which $\ao12$ representations  occur in a given type of $\lao22$ representations.  In Section \bref{sec:ads_grav} we explicitly consider how the spin-2 massless field theory  degresses from three to two dimensions. To this end we first consider a spin-$2$ massless field theory in AdS$_{d+1}$ spacetime (i.e. the linearized Einstein gravity with the cosmological constant) and then restrict to $d=2$ dimensions.  In Section \bref{sec:spin3} we extend the results and techniques of the previous section to the case of  spin-3 massless fields.  The conclusions and outlooks are given in Section \bref{sec:conc}. Appendix \bref{app:notation} summarizes our  notation and conventions for AdS spaces. In Appendix \bref{app:dec} we discuss the representation theory of  AdS$_2$ and AdS$_3$ isometry algebras in the context of adding discrete transformations inherited from the respective Lie groups. Appendix \bref{app:jacobi} contains an extended discussion of the Jacobi polynomials and associated basis functions needed to perform the \ldegr degression. Appendix \bref{app:rel} describes the Schouten identities in two and higher dimensions. All intermediate calculations from Section \bref{sec:spin3} are collected in Appendix \bref{app:spin3}.

\section{Summary  of the AdS$_3$/AdS$_2$ degression}
\label{sec:brief}

By analogy with the Poincare coordinates representing AdS$_{d+1}$ space as a stack of Minkowski spaces $\mathbb{R}^{d-1,1}$ with the growing warp factor, one may introduce new coordinates to represent AdS$_{d+1}$ as sliced by AdS$_d$ spaces of the growing radius which is effectively the $(d+1)$-th dimension  \cite{Metsaev:2000qb,Bak:2003jk}. Similar to the standard Kaluza-Klein analysis on a manifold being a direct product of two other manifolds, one considers a space $\cM^{d+1}$ sliced into AdS$_d$ hypersurfaces continuously parameterized by a finite interval variable with the full line element 
\be
\label{janus}
ds(\cM^{d+1})^2  = \frac{1}{\cos^2\theta}\left[ds(\text{AdS}_{d})^2 + \ell^2_{_{AdS}} d\theta^2\right]\,.
\ee      
Here, $\ell_{_{AdS}}$ is the AdS$_{d}$ radius and the slicing variable $\theta$ belongs to the closed interval $\theta \in[-\alpha, \alpha]$ with $\alpha$ restricted as  $\alpha < \pi/2$ \cite{Gwak:2016sma}. Note that parameters $\ell_{_{AdS}}$ and $\alpha$ define two independent scales. An asymptotic   $\alpha \to \pi/2$ defines  the decompactification limit analogous to the infinite radius limit of the Kaluza-Klein  manifold. In this limit $\cM^{d+1}$ goes back to AdS$_{d+1}$. The conformal boundary $\partial \text{AdS}_{d+1}$  consists of two limiting $\alpha = \pm \pi/2$ hypersurfaces (see \cite{Bak:2003jk} for more details). 
 
At $\alpha \neq \pi/2$ one may develop a sort of compactification of AdS$_{d+1}$ called the AdS waveguide compactification \cite{Gwak:2016sma}. This is strictly analogous to the original Kaluza-Klein compactification (see e.g. \cite{Hinterbichler:2013kwa} for recent discussion). In the decompactifying limit $\alpha = \pi/2$ one deals instead with a different mechanism called the dimensional degression \cite{Metsaev:2000qb,Artsukevich:2008vy} (see also \cite{Gross:2017aos}).

In practice, at $\alpha = \pi/2$ it is more convenient to work with another coordinate form of  the line element  \eqref{janus}  obtained by the  change $\tan \theta  = \sinh z$ with $z\in (-\infty, \infty)$ \cite{Artsukevich:2008vy}
\be
\label{av_metric}
ds(\text{AdS}_{d+1})^2  =  \cosh^2z\left[ ds(\text{AdS}_{d})^2  + \frac{\ell^2_{_{AdS}}}{\cosh^2z}dz^2  \right]\,,
\ee      
where we explicitly singled out the warp factor to identify the conformal boundary at $z = \pm\infty$. In the sequel we use the $d=2$ version of  \eqref{av_metric}.\footnote{A foliation of AdS$_3$ by AdS$_2$ slices was recently discussed in  \cite{Gutperle:2020gez}.}

\paragraph{Spin-$s$ massless fields.} Below we outline the basic ingredients of  the AdS$_3$/AdS$_2$ degression for spin-$s$ massless fields. The general approach is then illustrated by the examples of spins $s=2$ and $s=3$  in Sections \bref{sec:ads_grav} and \bref{sec:spin3}. More extended discussion of the higher-dimensional  AdS$_{d+1}$/AdS$_d$ degression  can be found in \cite{Metsaev:2000qb,Artsukevich:2008vy,Gwak:2016sma}.  

Let us consider the Fronsdal theory of double-traceless totally-symmetric tensor gauge  fields $\Phi^{m_1 ... m_s}(x,z)$ \cite{Fronsdal:1978rb,Lopatin:1988hz,Buchbinder:2001bs} which describe free massless spin-$s$ particles propagating in the AdS$_{d+1}$ space with the metric \eqref{av_metric} in the local AdS$_d$ coordinates $x$ and the slicing coordinate $z$. It is convenient to keep the dimension of space $d$ arbitrary. Of course, in $d+1 =3$ dimensions the Fronsdal theory becomes topological that is explicitly manifested within the Chern-Simons theory with the gauge algebra   $sl(N)\oplus sl(N)$ \cite{Blencowe:1988gj,Campoleoni:2010zq,Campoleoni:2011tn}.

A rank-$s$ tensor field can be  decomposed into $s+1$ lower rank component fields with convention  that Latin indices run $d+1$ values and Greek indices  run $d$ values:
\begin{equation}
\label{fronsdal}
\Phi^{m_1 ... m_s}(x,z) = \big\{\Phi^{\mu_1 ... \mu_s}(x,z),\; \Phi^{\mu_1 ... \mu_{s-1}}(x,z),\; \ldots, \Phi^{\mu_1}(x,z),\; \Phi(x,z)\big\}\,.
\end{equation}
The component fields here satisfy appropriate trace conditions followed from the Fronsdal double-traceless condition. 

\paragraph{Eigenfunction expansions.} In order to obtain a degressed  theory in one less dimension one integrates out the slicing  coordinate $z$.  This can be done by   expanding  the component functions \eqref{fronsdal}  with respect to  orthonormal sets of eigenfunctions (basis functions) $P^{t}_n(z)$ of auxiliary second-order differential operators in the $z$-coordinate. These operators   naturally arise in the original AdS$_{d+1}$ Fronsdal action written in terms of the metric \eqref{av_metric}. In this way, a rank-$t$ component field 
\begin{equation}
\label{phi_degr}
\Phi^{\mu_1 ... \mu_t}(x,z) = \sum^\infty_{n=0} \phi_n^{\mu_1 ... \mu_t}(x) \, P^{t}_n(z)\,,
\qquad t = 0,...,s\;,
\end{equation}
decomposes into an infinite collection of rank-$t$ totally-symmetric tensor  fields in AdS$_d$ space. Note that the basis functions $P^{t}_n(z)$ are labelled by $t$ which means that a rank-$t$ component field $\Phi^{\mu_1 ... \mu_t}(x,z)$ can have its own mode expansion. Remarkably, there is a unified choice of  basis functions  in terms of  the Jacobi polynomials $J_n^{\alpha,\beta}$ of running degrees for all rank-$t$ component fields, $P^{t}_n(z) \sim (\cosh z)^\gamma J_n^{\alpha,\beta}(-\tanh z)  $  (for some $\alpha, \beta, \gamma$ in terms of $d,n,t$; see  Appendix \bref{app:jacobi}). To integrate out the slicing coordinate $z$ in the original action one  calculates  the following overlaps of two basis functions
\begin{equation}
\label{overlap1}
\int_{-\infty}^{+\infty} dz \cosh^k z \,\left(A P^{t}_n\right)(z) \left(BP^{t^\prime}_m\right)(z)\;,
\end{equation}
for some integers $k, n, m, t, t^\prime$, and  some differential-algebraic operators  $A$ and $B$ acting on the basis functions. All ingredients in  the overlap integrals \eqref{overlap1} are constituents of the original second-order AdS$_{d+1}$ action written in the coordinates $x,z$ \eqref{av_metric}  so that the operators $A$ and $B$ are at most linear in the $z$-derivatives and involve hyperbolic coefficient functions.

\paragraph{Triangular transformations.}  In general, taking overlap integrals \eqref{overlap1} is a technically complicated problem.  It can be largely overcome by making triangular  field redefinitions of the components fields $\Phi^{\mu_1 ... \mu_t}(x,z)$. E.g. a rank-2 component can be redefined as 
\begin{equation}
\label{fldsbst}
\Phi^{\mu_1\mu_2}\;\; \rightarrow \;\;\widetilde{\Phi}^{\mu_1\mu_2}=\Phi^{\mu_1\mu_2}-\frac{1}{(d-2)} 
\text{sech}^{2} z\, g^{\mu_1 \mu_2} \Phi\,,
\end{equation}
where $g^{\mu_1 \mu_2}$ is the AdS$_d$ metric, see \eqref{eq:apanz}.  
Such a  redefinition  is typical  in the Kaluza-Klein type compactifications (see e.g. a review  \cite{Hinterbichler:2011tt}). 

Simultaneously, the above triangular transformations  diagonalize the bottom action that makes it possible to recognize the spectrum in a standard fashion as a sum of free field theories with fixed spins and masses. In particular, cancelling off-diagonal terms fixes the numerical coefficient in \eqref{fldsbst}. Strictly speaking, a diagonalization may also require partial gauge fixing.  On the other hand, the presence of a pole in the space dimension  does not allow making such a  substitution in the $d=2$ case.  A properly modified   diagonalization procedure  in two dimensions involves the use of particular basis functions different from those used in the higher-dimensional case as well as some extra identities to be discussed later in this section.  

\paragraph{Stueckelberg shift symmetry.}  Since  AdS$_{d+1}$ covariant derivatives contain algebraic contributions, then the gauge transformations of the component fields inherit such algebraic terms which are generally treated as Stueckelberg-type transformations. Indeed, the original Fronsdal field transforms as $\delta \Phi^{m_1 ... m_s}(x,z) = \ol\nabla^{\, (m_1}\Xi^{m_2...m_s)}(x,z)$, where $\Xi^{m_1...m_{s-1}}$ is a traceless totally-symmetric tensor gauge  parameter  and $\ol\nabla$ is the AdS$_{d+1}$ covariant derivative (see Appendix \bref{app:notation}). Similar to \eqref{fronsdal}  the gauge parameter decomposes as
\begin{equation}
\label{fronsdal1}
\Xi^{m_1 ... m_{s-1}}(x,z) = \big\{\xi^{\mu_1 ... \mu_{s-1}}(x,z),\; \xi^{\mu_1 ... \mu_{s-2}}(x,z),\; \ldots, \xi^{\mu_1}(x,z),\; \xi(x,z)\big\}\,,
\end{equation}
into $s$ independent gauge parameters subjected to their own trace conditions. Then, the component fields schematically transform as 
\be
\label{stuck}
\delta \Phi^{\mu_1 ... \mu_t}(x,z) = \nabla^{(\mu_1}\xi^{\mu_2...\mu_{s})}(x,z)+ \text{S}^{\mu_1 ... \mu_t}\big[\xi(x,z),g(x),\cosh z , \partial_z\big]\;,  
\ee 
where $t=0,...,s-1$ and S denotes a Stueckelberg-type  gauge operator which generally depends on   component gauge parameters,  AdS$_d$ metric, hyperbolic functions in $z$, and first derivatives in $z$. The basic rationale behind this formula is that the AdS$_d$ covariant derivative $\nabla$ encoded in the AdS$_{d+1}$ derivative $\ol\nabla$ is separated from other contributions which, therefore, contain the AdS$_d$ metric and are at most of first order in the $z$-variable.

The mode expansion for the component gauge parameters goes along the same lines as for the component fields \eqref{phi_degr} yielding infinite sets of gauge parameters $\xi^{\mu_1...\mu_t}_n(x)$ with $n=0,1,2,...,\infty$. It turns out that most of the field expansion modes $\phi_n^{\mu_1...\mu_t}$ are Stueckelberg fields which can be gauged away. The remaining fields define a bottom  theory: these  are generally massive if compared with the original Fronsdal massless AdS$_{d+1}$ field. The group-theoretical analysis in higher dimensions $d>2$ shows that there should arise  infinite collections of massive AdS$_d$ fields with  running masses and spins $0,1,...,s$ \cite{Artsukevich:2008vy,Gwak:2016sma}.              

\paragraph{Schouten identities and PDoF.} We now move on to considering the \ldegr degression. As discussed earlier, the bottom action in this case  cannot be directly   made diagonal so that there is a problem of identifying the spectrum.  
Nonetheless, only finitely many (off-)diagonal lower-rank terms remain non-vanishing on-shell while an infinite number of higher-rank contributions can be systematically neglected by using specific two-dimensional identities. As a by-product, an expected  infinite collection of massive fields on AdS$_2$ space reduces to just two massive spinless modes described by  the Klein-Gordon and Proca actions.\footnote{Note  that e.g. for gravitons  the \ldegr degression should be contrasted with the standard Kaluza-Klein compactification on $\mathbb{R}^{1,1}\times S^1$. The latter possesses the lowest eigenfunction with zero eigenvalue so that after the gauge fixing the remaining fields are massless. It follows that the two-dimensional bottom theory  has non-dynamical fields only  and, therefore, there are no PDoF \cite{Hinterbichler:2013kwa}. Contrary, in the AdS degression the lowest modes are massive and so that there are finitely many dynamical fields in the bottom theory  that changes the count of PDoF. We are grateful to K. Hinterbichler for a useful discussion of this issue.} 

Indeed,  AdS$_2$ kinetic  operators for rank-$s$  massive  fields  of the vacuum energy $E$ are of the form
\be
\label{ads2fron}
\left(-\Box + \ldots +m_s^2\right)\phi^{\mu_1 ... \mu_s}(x) \;, 
\quad
\text{where \cite{Metsaev:1997nj}:}
\qquad 
m_s^2 = E(E-1)-s\;,
\ee
where the ellipses denote second-order derivative and trace terms  which can be set to zero by imposing the TT (transverse + traceless) gauge conditions.  However, all these kinetic operators trivialize except for rank 0 and 1 fields described by the Klein-Gordon and Proca actions. In those cases we have the  field equations
\be
\label{KGMP}
\ba{l}
\text{AdS$_2$ Klein-Gordon theory:}\qquad\;\; \big[ -\Box + E(E-1)\big]\phi = 0\;,
\vspace{2mm}
\\
\text{AdS$_2$ Maxwell-Proca theory:}\qquad \big[-\Box + E(E-1)-1\big]\phi^\mu = 0\;, \quad \nabla_\mu \phi^\mu = 0\;.
\ea
\ee  
These two theories describe the same physical degrees of freedom spanned by the  $\ao12$ Verma module  of weight (energy)  $E$. This agrees  with our understanding that the only local degrees of freedom in AdS$_2$ are massive spinless modes. Both the Klein-Gordon and Proca theories with masses as those in \eqref{KGMP} have the same on-shell but differently realized off-shell dynamical content.

The above-mentioned vanishing of the higher-rank kinetic operators is due to the so-called  Schouten identities  which tell  us that the derivative part of a given kinetic operator is equivalent to a combination of algebraic terms (see Appendix \bref{app:rel}). So, roughly speaking, in the higher-spin sector of the AdS$_2$  theory almost all of infinite number of equations of motion  become algebraic. Basically, this means that  all fields $\phi_n^{\mu_1 ... \mu_s}(x)$ either vanish on-shell except for the zeroth scalar and vector  modes $\phi_0(x)$  and $\phi_0^\mu(x)$, or are expressed in terms of $\phi_0(x)$  and $\phi_0^\mu(x)$ (i.e. are auxiliary fields). 

To summarize, the spectrum of a bottom theory in the \ldegr degression  can be read off by means of the following multistage  procedure: {\bf(1)} specific basis functions, {\bf(2)} Stueckelberg-type gauge fixing, {\bf(3)} Schouten identities, {\bf(4)} eliminating  auxiliary fields. The resulting  sharp shortening of the physical spectra in two dimensions as compared to higher dimensions has a clear group-theoretical explanation given in the next section.

\section{Branching rules from  $\lao22$ to $\ao12$}
\label{sec:branch}

The global isometry of AdS$_{d+1}$ is o$(d,2)$ algebra and its lowest-weight (non-)unitary representations identified with elementary particles are  generalized Verma modules $\cD_{o(d,2)}(E,s)$  characterized by  energy $E$ and spin $s$ (here, we consider only totally-symmetric representations, see e.g. \cite{Nicolai:1984hb,deWit:1999ui}). At critical values $E_0 =  s+d-t-2$ defined by the  depth parameter  $t\in \{0,1,...,s-1\}$, there are singular submodules $\cS_t \subset \cD(E_0,s)$ generated from singular vectors on the $(t+1)$-th level. These are given by $\cS_t  = \cD(E_0', s')$, where $E'_0 = E_0+t+1$ and $s' = s - t - 1$. Factoring them out yields irreducible quotients
\be
\label{quotient}
\cH(E_0,s) = \cD(E_0,s)\big/\cS_t\;.
\ee
In AdS$_{d+1}$ space the resulting representations $\cH(E_0,s)$ describe  either non-unitary partially-massless spin-$s$ fields (depth $t>0$) \cite{Deser:2001pe} or unitary massless spin-$s$  fields ($t=0$) \cite{Fronsdal:1978rb}. Representations $\cD(E,s)$ with the energy above the unitary bound $E> E_0(t=0)$ describe unitary massive spin-$s$ fields.         

In lower dimensions $d=1$ and $d=2$ the above-described construction remains basically the same with a few additional specifications. See Appendix \bref{app:dec} for a detailed discussion.

\paragraph{$\ao12$ representations.} In $d=1$ all representations of the AdS$_2$ isometry algebra  $\ao12$ are characterized by a single parameter which can be interpreted as the energy $E\in \mathbb{R}$, while the spin number is absent. The respective lowest-weight representations are $\ao12$ Verma modules $\cD_{E}$. At $E>0$ the representations are unitary, irreducible, and infinite-dimensional. At negative values of energy $E\leq0$ the representations are no longer unitary. Moreover, when $E_0 = -j\in \frac{1}{2}\mathbb{Z}_{\hspace{-1mm}\leq\hspace{-0.8mm} 0}$, there is a singular submodule $\cS_{E_0}  = \cD_{-E_0+1}$ so that one may consider the quotient $\nD_{E_0} = \cD_{E_0}/\cS_{E_0}$ which is a finite-dimensional ($\dim \nD_{E_0}$ = $2j+1$) non-unitary $\ao12$ representation.\footnote{Except for $\nD_0$ which is unitarizable since it is a one-dimensional representation.}  

\paragraph{$\lao22$ representations.} In $d=2$ the AdS$_3$ isometry algebra is not simple and decomposes as $\lao22 \approx \ao12\oplus \ao12$. Parameterizing the energy and spin as $E = h_1+h_2\geq 0$ and $s = |h_1-h_2|$  one can see that the lowest-weight $\lao22$ spin-$s$ representations $\cD(E,s)$ decompose into $\ao12$ representations as
\be
\label{decom}
\cD(h_1+h_2,h_1-h_2) = \big[\cD_{h_1} \otimes \cD_{h_2}\big] \oplus  \big[\cD_{h_2} \otimes \cD_{h_1}\big]\;,
\ee
at $s\neq 0$ and 
\be
\label{decom0}
\cD^*(2h,0) = \cD_{h} \otimes \cD_{h}\;,
\ee
at $s=0$.  Each  $\big[\text{factor}\big]$ in \eqref{decom}  corresponds to modes $\pm s$ which form together  a parity-invariant combination. In the spin $s=0$ case the representation \eqref{decom0} is parity-invariant per se. In particular, it follows that $\cD(2h,0) = \cD^*(2h,0)\oplus \cD^*(2h,0)$.  Such representations are realized in $\lgo22$ invariant QFT, i.e. when the respective  space of states is invariant under $\lgo22$ discrete symmetries which are time reversal and space reflection,   see Appendix \bref{app:dec}.    

%(conversely, $h_1 =(E+s)/2$ and $h_2 = (E-s)/2$)

Since $\lao22$ representations under considerations are generally given by tensor products of $\ao12$ Verma modules, then they  can  be explicitly evaluated by means of the Clebsch-Gordan decompositions   (see Appendix \bref{app:dec}) 
\be
\label{CG}
\;\;\;\;\;\;\cD_{h_1} \otimes \cD_{h_2}\;\; =\;\;\;\; \bigoplus_{h=h_1+h_2}^{\infty}\, \cD_{h}\;,
\ee
\be
\label{CG0}
\;\;\;\;\;\;\cD_{h_3} \otimes \nD_{h_4} \;\; =\;\;\;\; \bigoplus_{h=h_3+h_4}^{h_3-h_4}\, \cD_{h}\;,
\ee
where $\forall h_1, h_2$ and $h_3>0$, $h_3 > \abs{h_4}$, $h_4\in \frac{1}{2}\mathbb{Z}_{\hspace{-1mm}\leq\hspace{-0.8mm} 0}$, summation in $n$ goes with a step $1$. 
 
\paragraph{Branching rules for massive representations.}   Substituting \eqref{CG} in \eqref{decom}  we  obtain 
\be
\label{branch_mas}
\cD(E,s)\; = \; \bigoplus_{n=0}^\infty  \big[\cD_{E+n}\oplus \cD_{E+n}\big]\; \equiv  \;2\, \bigoplus_{n=0}^\infty  \cD_{E+n}\;,
\ee
where the factor of 2 indicates that each representation  is duplicated. The scalar representation \eqref{decom0} decomposes as
\be
\label{branch_scalar}
\cD(E,0)\; =\; \bigoplus_{n=0}^\infty  \cD_{E+n}\;,
\ee
that precisely matches the branching rule for a scalar $o(d,2)$ representation   \cite{Artsukevich:2008vy} evaluated at $d=2$. The above branching rules show that a given $\lao22$ generalized Verma module  decomposes into an infinite collection of $\ao12$ Verma modules with an equidistant spectrum of weights.

\paragraph{Branching rules for (partially-)massless representations.} In order to formulate  the branching rules for the quotient representations \eqref{quotient} in $d=2$ dimensions  we  use the Clebsch-Gordan series \eqref{CG} and obtain both  the original representation and the singular submodule in the form   
\be
\label{branch_gen}
\cD(s-t,s)\; =\; 2\bigoplus_{n=0}^{\infty} \cD_{s-t+n}  \equiv \big[2\, \bigoplus_{k=0}^{t}\cD_{s-k}\big]\oplus  \big[ 2\, \bigoplus_{m=1}^{\infty} \cD_{s+m} \big]
\ee 
\be
\label{singular}
\cS_t = \cD(s+1,s-t-1) = 2\, \bigoplus_{m=1}^{\infty} \cD_{s+m}\;.
\ee
Here, a summation in  \eqref{branch_gen} is reorganized to isolate  the singular submodule \eqref{singular}. Taking the quotient \eqref{branch_gen}/\eqref{singular} we are left with the first factor in \eqref{branch_gen} given by 
\be
\label{branch_D}
\cH(s-t,s) =  \bigoplus_{n=0}^{t}\big[\cD_{s-n}\oplus \cD_{s-n}  \big]\;.
\ee
This is the branching rule for the (partially-)massless $\lao22$ representations. We find out  that, contrary to infinite towers of representations in $d\geq 3$, we have  just a finite number of representations in $d=2$.  

According to \cite{Gwak:2015jdo,Basile:2018dzi}  the maximal-depth case $t=s-1$ has a degenerate interpretation. The corresponding singular submodule \eqref{singular} is given by the duplicated scalar representation  $\cD^*(s+1,0)\oplus \cD^*(s+1,0)$, where each factor is  parity-invariant on its own. In principle, one can consider a quotient representation with either  one or two scalar submodules factored out. Factoring out two copies  $\cD^*(s+1,0)\oplus \cD^*(s+1,0)$ we obtain \eqref{branch_D}, while factoring out only one copy  $\cD^*(s+1,0)$ yields 
\be
\label{partial_II}
\tilde\cH(s-t,s) = \cH(s-t,s)\oplus\cD^*(s+1,0) \equiv \bigoplus_{n=0}^{t}\big[\cD_{s-n}\oplus \cD_{s-n}  \big] \bigoplus_{n=0}^\infty  \cD_{s+n+1}\;.
\ee    

On the other hand, the right-hand side of the branching rule \eqref{branch_D} can be packed again into the tensor product by virtue of the Clebsch-Gordan series  \eqref{CG0} to obtain \cite{Gwak:2015jdo} 
\be
\label{branch_D1}
\cH(s-t,s) = \big[\cD_{s-\frac{t}{2}}\otimes \nD_{-\frac{t}{2}}\big]\oplus \big[\nD_{-\frac{t}{2}}\otimes \cD_{s-\frac{t}{2}}\big] \;,
\ee
where $\nD_{-\frac{t}{2}}$ is a $(t+1)$-dimensional $\ao12$ representation. In particular, \eqref{branch_D1} explicitly shows that all $t\neq 0$  representations are non-unitary due to $\nD_{-\frac{t}{2}}$ factor which is necessarily non-unitary being finite-dimensional. The unitarity holds in the case  $t=0$ only.

\paragraph{Summary.} Relations \eqref{branch_mas}, \eqref{branch_scalar}, \eqref{branch_D}, and \eqref{partial_II} are the branching rules for, respectively,  massive spin-$s$, massive(massless) spin-$0$, and (partially-)massless spin-$s$ depth-$t$ $\lao22$ representations. Below is the summary list of these representations. 
\be
\label{branch_list}
\ba{l}
\vspace{-2mm}

\dps
\text{massive(massless) spin-$0$:}\hspace{12mm}\cD(E,0) = \bigoplus_{n=0}^\infty\,  \cD_{E+n}
\\
\\
\dps
\text{massive spin-$s$:}\qquad\hspace{21.5mm}\cD(E,s) = \bigoplus_{n=0}^\infty  \big[\cD_{E+n}\oplus \cD_{E+n}\big]
\\
\\
\vspace{2mm}
\dps
\text{massless spin-$s$:}\qquad\hspace{21.0mm}\cH(s,s) = \cD_{s}\oplus \cD_{s}
\\
\\
\dps
\text{partially-massless spin-$s$ (I):}\qquad\cH(s-t,s) = \big[\cD_{s-t}\oplus \cdots \oplus \cD_{s}\big]\oplus \big[\cD_{s-t}\oplus \cdots \oplus \cD_{s}\big]
\vspace{-1.7mm}
\\
\text{any depth}
\\
\\
\dps
\text{partially-massless spin-$s$ (II):}\hspace{7mm}\tilde\cH(1,s) = \cH(1,s) \bigoplus_{n=0}^\infty  \cD_{s+n+1}
\vspace{-4.7mm}
\\
\text{maximal depth}

\ea
\ee

In the next sections we consider the   field-theoretical realization of the massless spin-$s$ representations $\cH(s,s)$. In particular, from our discussion in Section \bref{sec:brief} we conclude that  regardless of spin value $\cH(s,s)$  are described  by the Klein-Gordon and Proca equations in AdS$_2$ space  which correspond to two factors on the right-hand side of the respective  branching rule. Note that the maximal depth partially-massless systems can have or not have  local degrees of freedom depending on a local field theory formulation. In particular, the Maxwell theory is simultaneously a massless spin-1 and a maximal depth  partially-massless spin-1 system which is classified in the list  \eqref{branch_list} by the last line. It follows that the \ldegr degression of the Maxwell theory yields an infinite spectrum. Thus, the phenomenon of finite spectra begins to  fully manifest itself when $s=2$.

\section{AdS$_3$/AdS$_2$ degression of spin-2 massless fields}
\label{sec:ads_grav}

Consider first  the linearized action of  $(d+1)$-dimensional gravity with the cosmological term. The resulting  theory of symmetric traceful rank-2 tensor field  $\Phi^{mn} = \Phi^{mn}(x,z)$  describes a massless spin-2 particle propagating in AdS$_{d+1}$ spacetime,
\begin{al}
\label{eq:s2act}
S= \int d \mu_{d+1}\Big\{& - (\ol{\nabla}_{a}  \Phi^{mn})^2 +2 (\ol{\nabla}_{m}  \Phi^{mn})^2  -2 \ol{\nabla}_{m}  \Phi \ol{\nabla}_{n}  \Phi^{mn} + (\ol{\nabla}_{a}  \Phi)^2 +  \\
& + b\left( 2( \Phi^{mn})^2 + (d-2)\Phi^2 \right) \Big\}\,, \qquad  \Phi\equiv  \Phi^{mn}G_{mn}\,,
\end{al}%
where for future convenience we introduced the concise notation for the integration measure $d\mu_{d+1}$  \eqref{int_meas_1} and $G_{mn}(x,z)$ is the AdS$_{d+1}$ metric \eqref{eq:apanz} (see Appendix \bref{app:notation} for other conventions and notation). The action is invariant under the gauge transformations with the gauge parameter $\Xi^m$ 
\begin{equation}\label{eq:s2gau}
\delta  \Phi^{mn}(x,z) = \ol{\nabla}^{m} \Xi^{n}(x,z)+\ol{\nabla}^{n} \Xi^{m}(x,z)\,,
\end{equation}
where $\ol{\nabla}$ stands for the AdS$_{d+1}$ covariant derivative. 
Throughout this section we keep the dimension of space $d$ arbitrary and  set $d=2$ in the concluding subsection \bref{sec:2d_deg}.

\subsection{$d+1$ split and component actions}

According to the general strategy described in Section \bref{sec:brief} we decompose  a rank-2 field  $\Phi^{mn}$ into rank-$0, 1, 2$ component fields  as 
\begin{equation}
\label{components3}
\Phi^{mn}(x,z) = \big\{ \Phi^{\mu\nu}(x,z),\; \Phi^{\mu}(x,z),\; \Phi(x,z)\big\} \coloneqq \big\{ h^{\mu\nu}(x,z),\; A^\mu(x,z),\; \phi(x,z)\big\}\,,
\end{equation}
along with the gauge parameter 
\begin{equation}
\label{components3_g}
\Xi^{m}(x,z) = \big\{\Xi^\mu(x,z),\; \Xi(x,z) \big\} \coloneqq \big\{ab^{-1}\cosh^2z\,\xi^\mu(x,z),\;ab^{-1}\cosh^2z\,\xi(x,z) \big\} \,,
\end{equation}
where for future convenience we renamed and  redefined component tensors by means of the hyperbolic functions, and $a,b$ are convenient numerical factors parameterizing AdS curvatures (see Appendix \bref{app:notation}).  Then, plugging \eqref{components3} and \eqref{components3_g} as well as the metric $G_{mn}$ \eqref{eq:apanz} into the action \eqref{eq:s2act} we find the component representation  
\begin{equation}
 S = \sum_{m\geq n}\,\sum_{n=0,1,2}S_{mn}\,,
\end{equation} 
where  
\begin{align}
S_{22} =& a\iint  \cosh^2z\Big\{ -(\nabla_\alpha h^{\mu\nu})^2 + 2(\nabla_\mu h^{\mu\nu})^2 - 2\nabla_\mu h\nabla_\nu h^{\mu\nu} + (\nabla_\alpha h)^2 + \label{AV_new22}
 \\
   &+ a\big[ 2(h^{\mu\nu})^2 + (d-3)h^2 + h_{\mu\nu}\,\LL_d\,(\cosh^2z\,\LL_2\,h^{\mu\nu}) - h\LL_d(\cosh^2z\,\LL_2\,h) \big]\Big\}\,,\notag\\
 S_{21} =&  4a\iint \cosh^2z\Big\{ \nabla_\mu h^{\mu\nu}\,\LL_d \,A_\nu - \nabla_\mu h \, \LL_d \, A^\mu\Big\}\,,\\
 S_{20} =&  2\iint \Big\{ \phi\big[\nabla_\mu\nabla_\nu h^{\mu\nu} - \Box h\big] + a(d-1)\phi h - a(d-1)\phi \tanh z\cosh^2z\, \LL_2\,h\Big\}\,,\\
 S_{11} =&  2\iint \Big\{ -(\nabla_\mu A^\nu)^2 + (\nabla_\mu A^\mu)^2 - a(d-1)(A^\mu)^2 \Big\}\,,\\
 S_{10} =&  4(d-1)\iint \tanh z\,\nabla_\mu A^\mu \phi\,,\\ 
 S_{00} =&  d(d-1)\iint  \tanh^2 z\,\phi^2  \label{AV_new00}\,.
\end{align}
Here, $h=g_{\mu\nu}h^{\mu\nu}$ and $\Box = \nabla_\mu\nabla^\mu$, also we   introduced the  double-integral notation  \eqref{int_meas_1} 
\begin{equation}
\label{measure2}
\iint=b^{-1} \int d \mu_{d+1}=a^{d/2}b^{-(d+3)/2} \int d \mu_{d} \int d z\, \cosh ^{d} z \,.
\end{equation}
The component actions \eqref{AV_new22}-\eqref{AV_new00} are essentially  those obtained in \cite{Artsukevich:2008vy}. Here we slightly rearranged terms and denoted different action components as $S_{mn}$ to indicate contributions of rank-$m$ and rank-$n$ components. 
Also, following \cite{Gwak:2016sma} we introduced a convenient notation $\LL_m$ \eqref{eq:apL} for particular  first-order differential operators in the $z$-coordinate.   

The gauge transformations \eqref{eq:s2gau} are given by  
\begin{align}
&\delta h^{\mu \nu} =\nabla^{\mu} \xi^{\nu}+\nabla^{\nu} \xi^{\mu} + 2\tanh{z}\, g^{\mu \nu}\xi \,, \label{eq:s2gauz1}\\
&\delta A^{\mu} = \nabla^{\mu} \xi + a\cosh^2{\hspace{-1mm}z} \,\LL_2 \xi^{\mu}\,,\label{eq:s2gauz2}\\
&\delta \phi =2a\cosh^2{\hspace{-1mm}z}\, \LL_2 \xi\,,\label{eq:s2gauz3}
\end{align}
according to the general form of the Stueckelberg-type transformations \eqref{stuck}.

\subsection{Integrating out the slicing coordinate}

The mode expansions for fields and parameters  read
\be
\label{mode_exp_f}
h^{\mu\nu}(x,z)= \sum_{n=0}^{\infty} h^{\mu\nu}_n(x)P^2_n(z)\,, 
\;
A^{\mu}(x,z) = \sum_{n=0}^{\infty} A_{n}^{\mu}(x) P_{n}^{1}(z)\,,
\;
\phi(x,z)= \sum_{n=0}^{\infty} \phi_{n}(x) P_{n}^{1}(z)\,,
\ee
\be
\label{mode_exp_g}
\xi^{\mu}(x,z) = \sum_{n=0}^{\infty} \xi_{n}^{\mu}(x) P_{n}^{2}(z)\,,
\qquad
\xi(x,z)= \sum_{n=0}^{\infty} \xi_{n}(x) P_{n}^{2}(z)\,,
\ee
where the basis functions $P^s_n$  are related to the Jacobi polynomials (see Appendix \bref{app:basis} for exact expressions). There is an essential difference between $d=2$ and $d> 2$ cases shortly discussed in Section \bref{sec:brief}. Namely, making  the field redefinition  \eqref{fldsbst}  in $d > 2$ one can see that the most convenient mode expansion for the scalar component is given by \cite{Artsukevich:2008vy} 
\be
\phi(x,z) = \sum_{n=0}^{\infty} \phi_{n}(x) P_{n}^{0}(z)\;,
\ee 
where the basis functions $P_n^0$ are used instead of $P_n^1$.  
In this case the overlap integrals possess the  properties \eqref{eq:PPdelta} and \eqref{LPmP} which  are sufficient to integrate out the $z$-coordinate in a straightforward manner. However,  it turns out that the basis functions $P^0_n$ do not exist in $d=2$ as their inner products do not converge (see the beginning of Appendix \bref{app:basis}). This fact along with the $d=2$ pole in the substitution \eqref{fldsbst} lead  us to fix  in the mode expansions \eqref{mode_exp_f} the same basis functions both for scalar and vector components. It follows  the resulting inner products between basis functions are much more complicated than those in $d\neq 2$ dimensions. However, using the calculation technique elaborated  in  Appendix \bref{app:inn} all relevant inner products can be explicitly evaluated in analytical  terms. On the other hand, the choice of  basis functions for the component gauge parameters as in \eqref{mode_exp_g}  greatly simplifies the Stueckelberg-type operator \eqref{stuck} after the mode expansion, see \eqref{stuck_3} below. Of course, one can equally choose other  basis functions, but practical  computations show that the mode expansions \eqref{mode_exp_f} are optimal against other possible choices.

Now, inserting the mode expansions \eqref{mode_exp_f} into the component actions \eqref{AV_new22}-\eqref{AV_new00} and introducing the notation
\begin{equation}
\int \equiv a^{d / 2}b^{-(d+3) / 2} \int d \mu_{d} \,,
\end{equation}
we obtain 
\begin{align}
 S_{22} =& a\int\sum^{\infty}_{n=0}  \Big\{ -(\nabla_\alpha h_n^{\mu\nu})^2 + 2(\nabla_{\mu} h_n^{\mu\nu})^2 - 2\nabla_\mu h_n\nabla_\nu h_n^{\mu\nu} + (\nabla_\alpha h_n)^2 + \label{Ya_22}\\
  &+ a\Big[ (-(\mm{n})^2+2)(h_n^{\mu\nu})^2 + ((\mm{n})^2+d-3)h_n^2 \Big]\Big\}\,,\notag\\
 S_{21} =& 4a\int\sum^{\infty}_{n=1}\mm{n-1} \Big\{ -\nabla_\mu h_{n-1}^{\mu\nu}A_{n\nu} + \nabla_\mu h_{n-1} A_n^\mu \Big\}\,,\\
 S_{20} =& 2\int\sum^{\infty}_{n,m=0} \Big\{ \phi_n\big[\nabla_\mu\nabla_\nu h_m^{\mu\nu} - \Box h_m + a(d-1)h_m \big](P^1_n,P^2_m)_1 -\\
  & - a(d-1)\mm{m}\phi_n h_m(P^1_n,\tanh z P^1_{m+1})_1\Big\}\,,\notag\\
 S_{11} =& 2\int\sum^{\infty}_{n=0} \Big\{ -(\nabla_\mu A_n^\nu)^2 + (\nabla_\mu A_n^\mu)^2 - a(d-1)(A_n^\mu)^2 \Big\}\,,\\
 S_{10} =& 4(d-1)\int\sum^{\infty}_{n,m=0} \nabla_\mu A_n^\mu\, \phi_m\,(P^1_n,\tanh zP^1_m)_1 \,,\\ 
S_{00} =& d(d-1)\int\sum^{\infty}_{n,m=0}  \phi_n\phi_m\,(P^1_n,\tanh^2 zP^1_m)_1  \label{Ya_00}\,,
\end{align}
where the constant $\mm{n}$ is given by 
\be
\label{m2}
\mm{n} = \sqrt{(n+1)(n+d)}\;,
\ee
(a general definition is given in   \eqref{eq:m}). The inner products $(A,B)_l$ here  are defined in \eqref{eq:PPdelta}. 

 The gauge transformations \eqref{eq:s2gauz1}-\eqref{eq:s2gauz3} are  similarly expanded as   
\be
\label{stuck_1}
\delta h_n^{\mu \nu} =\nabla^{\mu} \xi_n^{\nu}+\nabla^{\nu} \xi_n^{\mu} + 2g^{\mu \nu}\sum^{\infty}_{m=0}(P^2_n,\tanh zP^1_m)_2\,\xi_m\,, 
\ee
\be
\label{stuck_3}
\delta A^{\mu}_n = \nabla^{\mu} \xi_n + a\mm{n-1}\, \xi_{n-1}^{\mu}\,,
\qquad
\delta \phi_n =2a \mm{n-1}\, \xi_{n-1}\,,
\ee
where $n=0,1, ..., \infty$. As a consistency check, one can directly verify  that the total  component action \eqref{Ya_22}-\eqref{Ya_00} is invariant under the mode transformations \eqref{stuck_1}-\eqref{stuck_3}. Their Stueckelberg-type form  suggests that the parameters $\xi^\mu_{m}$ and $\xi_{n}$ with $m,n = 0,1,...,\infty$ can be used to gauge away the fields  $A^\mu_{m}$ and $\phi_n$ with $m,n = 1,2,..., \infty$.  It follows that no  gauge parameters and residual gauge symmetry are left for the other  fields  $A^{\mu}_0$, $\phi_0$, and $h_n^{\mu \nu}$ with $n=0,1,..., \infty$. The corresponding total action $S$ is drastically simplified. It takes the form
\begin{align}
 S_{22} =& a\sum^{\infty}_{n=0}\int  \Big\{ -(\nabla_\alpha h_n^{\mu\nu})^2 + 2(\nabla_{\mu} h_n^{\mu\nu})^2 - 2\nabla_\mu h_n\nabla_\nu h_n^{\mu\nu} + (\nabla_\alpha h_n)^2 + \label{Yan_22}\\
  &+ a\Big[ (-(\mm{n})^2+2)(h_n^{\mu\nu})^2 + ((\mm{n})^2+d-3)h_n^2 \Big]\Big\}\,,\notag\\
 S_{21} =&0\,,\\
 S_{20} =& 2\int \Big\{ \phi_0 \big[\nabla_\mu\nabla_\nu h_0^{\mu\nu} - \Box h_0 + a(d-1)h_0 \big](P^1_0,P^2_0)_1 -\label{Yan_20}\\
  & - a(d-1)\mm{0}\,\phi_0 h_0\,(P^1_0,\tanh z P^1_{1})_1\Big\}\,,\notag\\
 S_{11} =& 2\int \Big\{ -(\nabla_\alpha A_0^\mu)^2 + (\nabla_\mu A_0^\mu)^2 - a(d-1)(A_0^\mu)^2 \Big\}\,,\\
 S_{10} =&0\,,\\ 
 S_{00} =& d(d-1)\int \Big\{ \phi_0^2\,(P^1_0,\tanh^2 zP^1_0)_1 \Big\}\label{Yan_00}\,,
\end{align}
where the inner products $(P^1_0,P^2_0)_1$, $(P^1_0,\tanh z P^1_{1})_1$, $(P^1_0,\tanh^2 zP^1_0)_1$ are  constants (depending on $d$) which can be calculated using the technique elaborated in Appendix \bref{app:inn}. 

At this stage, despite the use of different sets of the basis functions we observe that the resulting set of fields does  coincide with that one  found in \cite{Artsukevich:2008vy}: an infinite tower of rank-2 fields, a single vector field, a single scalar field. However, the action above contains off-diagonal terms and a priori it is not clear if  the spectrum of masses and spins is correct. In particular, from the component actions  \eqref{Yan_20} and \eqref{Yan_00} involving  a scalar field $\phi_0$ it follows that $\phi_0$ misses a standard kinetic term. It is a triangular algebraic  field redefinition \eqref{fldsbst} that introduces a required kinetic term for $\phi_0$. Being expanded in modes the triangular redefinition formula   goes like $\tilde h_n^{\mu\nu} =  h_n^{\mu\nu}+ g^{\mu\nu}\sum_{m}T_{mn} \phi_m$, where $T_{mn}$ are some overlap coefficients which can be explicitly calculated. In order to diagonalize the action more field redefinitions are required which are not explicitly seen for the above choice of the basis functions.

\subsection{Degressed equations of motion in two dimensions}
\label{sec:2d_deg}

Let us finally fix $d=2$. Then, various constants arising in the higher dimensional action  can be explicitly calculated  as  
\begin{equation}
\mm{0} = \sqrt{2}\,, \quad (P^1_0,P^2_0)_1 = \sqrt{\frac{2}{3}}\,, \quad (P^1_0,\tanh z P^1_{1})_1 = -\frac{1}{\sqrt{3}}\,,\quad (P^1_0,\tanh^2 zP^1_0)_1 = \frac{1}{3}\,.
\end{equation}
The equations of motion that follow from the total action \eqref{Yan_22}-\eqref{Yan_00} at $d=2$ are given by 
\begin{align}
&\frac{\delta S}{\delta h^{\mu\nu}_n}=0: && n=0: \quad \Box h^{\mu\nu}_0 - \nabla^{(\mu}\nabla_\rho h^{\nu)\rho}_0  + \nabla^\mu\nabla^\nu h_0 + g^{\mu\nu}\nabla_\rho\nabla_\sigma h^{\rho\sigma}_0 -g^{\mu\nu}\Box h_0 + \label{eq:eom2}\\
&&&\qquad\qquad +ag^{\mu\nu}h_0 + \nabla^\mu\nabla^\nu\phi_0 - g^{\mu\nu}\Box\phi_0 + 2ag^{\mu\nu}\phi_0 = 0\,,\notag\\
&&&n\geq 1: \quad \Box h^{\mu\nu}_n - \nabla^{(\mu}\nabla_\rho h^{\nu)\rho}_n  + \nabla^\mu\nabla^\nu h_n + g^{\mu\nu}\nabla_\rho\nabla_\sigma h^{\rho\sigma}_n -g^{\mu\nu}\Box h_n + \label{eq:eom1}\\
&&&\qquad\qquad+a\big[ \big( -(\mm{n})^2+2 \big)h^{\mu\nu}_n+\big( (\mm{n})^2-1\big)g^{\mu\nu}h_n \big] = 0\,, \notag\\ 
&\frac{\delta S}{\delta A^\mu_0}=0: && \hspace{17mm}\Box A^\mu_0 - \nabla^\mu \nabla_\nu A^\nu_0 - aA^\mu_0 = 0\,,\label{eq:eom3}\\
&\,\frac{\delta S}{\delta\phi_0}=0 : && \hspace{17mm}\nabla_\mu \nabla_\nu h^{\mu\nu}_0 - \Box h_0 + 2ah_0 + a\phi_0 = 0\,.\label{eq:eom4}
\end{align}
To simplify the resulting expressions here we redefined the scalar field as ${a}^{-1}\sqrt{2/3}\phi_0 \to\phi_0$. The equation \eqref{eq:eom3} describes a massive vector field $A^\mu_0$ which decouples from  other fields. The other three equations \eqref{eq:eom2}, \eqref{eq:eom1}, \eqref{eq:eom4} form a coupled system of relations on the tower of rank-2 fields $h^{\mu\nu}_n$ and a scalar field $\phi_0$.  

Now, the Schouten identity  \eqref{schouten} for rank-2 tensor fields allows eliminating  all derivative  terms in \eqref{eq:eom2} and \eqref{eq:eom1}. Indeed, all derivative terms in the first lines of \eqref{eq:eom2} and \eqref{eq:eom1} (note that the first lines are identical) constitute the derivative part of the Schouten identity \eqref{schouten} so that they are equal to particular combination of purely algebraic terms. Thus, these equations  take much more elegant form
\begin{align}
& n=0: \quad && \nabla^\mu\nabla^\nu\phi_0 - g^{\mu\nu}\Box\phi_0 + 2ag^{\mu\nu}\phi_0 +2a\big[ -h^{\mu\nu}_0 + g^{\mu\nu}h_0 \big]= 0\;,\label{eq:s2eom2}\\
&n\geq 1: && (\mm{n})^2\big[ -h^{\mu\nu}_n + g^{\mu\nu}h_n \big] = 0\;.\label{eq:s2eom1}
\end{align} 
Since the parameter  $(\mm{n})^2 \neq 0$ for any $n$ \eqref{m2}, then the second equation   \eqref{eq:s2eom1} can be easily solved. Taking the trace results in $h_n = 0$ and, hence, $h^{\mu\nu}_n = 0$ at $n= 1,2,..., \infty$. Thus, these modes do not propagate. Taking the trace  in the first  equation \eqref{eq:s2eom2} yields 
\begin{equation}
\label{aux1}
h_0 = \frac{1}{2a}(\Box - 4a)\phi_0\,.
\end{equation}
Substituting this expression back into \eqref{eq:s2eom2}  we can express $h^{\mu\nu}_0$ in terms of the scalar field   as 
\begin{equation}
\label{aux2}
h^{\mu\nu}_0 = \frac{1}{2a}\big( \nabla^\mu\nabla^\nu - 2ag^{\mu\nu} \big)\phi_0\,.
\end{equation}
In other words, the rank-2 tensor field is auxiliary.  Finally, substituting  \eqref{aux1}-\eqref{aux2} into the last equation of motion \eqref{eq:eom4} we obtain the following  fourth-order equation 
\begin{equation}
\frac{1}{2a}\big[\nabla_\mu\nabla_\nu\nabla^\mu\nabla^\nu - 2a\Box - (\Box - 2a)(\Box - 4a)  \big]\phi_0 + a\phi_0 = 0\,.
 \end{equation}
However, one can see that the terms quartic in covariant derivatives cancel each other so that the resulting equation takes the standard second-order form
\begin{equation}
(\Box - 2a)\phi_0 = 0\,.
\end{equation}

To summarize, we obtain the infinite sequences of modes  vanishing on-shell  
\be
\label{res_1}
h_{n}^{\mu\nu} = 0\,,
\qquad
A_{n}^{\mu} = 0\,,
\qquad
\phi_{n} = 0\,,
\qquad 
n=1,2,...,\infty\;,
\ee
along with the zeroth rank-2 mode which is an auxiliary field
\be
h^{\mu\nu}_0 = \frac{1}{2a}\left[\nabla^\mu \nabla^\nu - 2ag^{\mu\nu}\right]\phi_0\,.
\ee
The remaining fields  are the scalar and vector zeroth modes subjected to the second-order equations
\be
\label{res_4}
\ba{l}
\dps
\big[-\Box + 2a\big]\phi_0 = 0\,,
\vspace{2mm}
\\
\big[-\Box + a\big]A^\mu_0 = 0\,,
\quad
\nabla_\mu A^\mu_0 = 0\,,
\ea
\ee
which are the Klein-Gordon and  Proca  equations, respectively. 
Fixing $a=1$ we conclude from \eqref{KGMP} that $E=2$ and PDoF are organized into two Verma modules $\cD_{2}\oplus \cD_{2}$ according to the branching rules \eqref{branch_list} for $s=2$.

To conclude this section, from \eqref{eq:eom4} we notice that the scalar field $\phi_0$ can  equivalently be treated as auxiliary with respect to the rank-2 field, i.e. $\phi_0 = a^{-1}(\nabla_\mu \nabla_\nu h^{\mu\nu}_0 - \Box h_0 + 2ah_0)$. Substituting this expression into \eqref{eq:eom3} we arrive at the forth-order equation on $h_0^{\mu\nu}$. In this form the final system becomes diagonalized at the cost of having higher-order equations. However, our previous consideration explicitly shows that the resulting higher-order equation is essentially of second-order by means of introducing an additional scalar variable. This phenomenon of a higher-derivative theory describing  unitary PDoF is similar to that of the NMG theory in three dimensions \cite{Bergshoeff:2009hq}.

\section{AdS$_3$/AdS$_2$ degression of spin-3 massless fields}
\label{sec:spin3}

A  massless spin-3  particle  in $AdS_{d+1}$  is described by a symmetric traceful tensor  $\Phi^{mnk}$ with the following gauge transformation\footnote{The AdS$_{d+1}$ waveguide compactification for  massless spin-3 particles was considered in  \cite{Gwak:2016sma}.}   
\begin{equation}
\label{gauge33}
\delta \Phi^{mnk} = \ol\nabla^{m}\Xi^{nk}+\ol\nabla^{n}\Xi^{mk}+\ol\nabla^{k}\Xi^{mn}\,,
\end{equation}
where the gauge parameter $\Xi^{mn}$ is a symmetric traceless tensor. Again, we begin in arbitrary dimension, then  specify to dimension two. The action is given by 
\be
\ba{l}
\dps
S = \int d\mu_{d+1}\Big\{ -(\ol\nabla_a \Phi^{mnk})^2+3(\ol\nabla_m \Phi^{mnk})^2 - 6\ol\nabla_m \Phi_n\ol\nabla_k \Phi^{mnk} + 3(\ol\nabla_a \Phi^m)^2  + 
\vspace{2mm}
\\
\dps
\hspace{20mm}+ \frac{3}{2}(\ol\nabla_m \Phi^m)^2 + b\big[ -(d-2)(\Phi^{mnk})^2 + 6d\,(\Phi^m)^2   \big]  \Big\} \,,
\qquad  \Phi^{m} \equiv \Phi^{mnk}G_{nk}\,.
\ea
\ee
Decomposing the original field as $\Phi^{mnk}(x,z) = \big\{\Phi^{\mu\nu\rho}(x,z),\, \Phi^{\mu\nu}(x,z),\,\Phi^{\mu}(x,z),\, \Phi(x,z)\big\}$ we introduce new notations for the component fields along with some convenient redefinitions
\be
w^{\mu\nu\rho} \coloneqq \Phi^{\mu\nu\rho} + (ad)^{-1}\sech^2 z\, g^{(\mu\nu}\Phi^{\rho)}\;, 
\quad
h^{\mu\nu} \coloneqq \Phi^{\mu\nu}\;, 
\quad
A^{\mu} \coloneqq \Phi^{\mu}\;, 
\quad
\phi \coloneqq \Phi\;.
\ee
For the component fields we choose the following mode expansions
\be
\label{modes_3}
\ba{c}
\dps
w^{\mu\nu\rho}(x,z) = \sum^\infty_{n=0} w^{\mu\nu\rho}_n(x)P^3_n(z)\,, 
\qquad
h^{\mu\nu}(x,z)= \sum^\infty_{n=0} h^{\mu\nu}_n(x)P^2_n(z)\,,
\vspace{1mm}
\\
\dps
A^\mu(x,z) = \sum^\infty_{n=0} A^{\mu}_n(x)P^1_n(z)\,, 
\qquad
\phi(x,z) = \sum^\infty_{n=0} \phi_n(x)P^1_n(z)\,.
\ea
\ee
Similar to the spin-2 case the basis functions of the scalar and vector components   are chosen to be the same since  $P^0_n$ do not exist in $d=2$ (see Appendix \bref{app:basis}) so  that $P^1_n$ are used  instead. 

The \hdegr degression for  massless spin-3  fields goes along the same lines as for  massless spin-2 fields in Section \bref{sec:ads_grav}. Therefore, all  the details of calculating  the degressed  component actions are relegated to Appendix \bref{app:spin3}. Here, we will assume that the Stueckelberg-type gauge symmetry with parameters $\xi^\mu_n$ and $\xi_n$ has been partially used to gauge away fields $A^\mu_n$ and $\phi_n$ with $n=1,2,...,\infty$ and the resulting  degressed equations of motion in AdS$_d$ are explicitly known, see eqs. \eqref{e1}-\eqref{e5}. Now, we fix $d=2$.

As shown in Section \bref{sec:ads_grav} the Schouten identities \eqref{schouten} for rank-2 tensor fields  are sufficient to turn the spin-2 kinetic operators  into purely algebraic terms that results in truncating the spectrum. However,  the Schouten identities  \eqref{Schouten3} for rank-3 tensor fields   are generally different from the spin-3 kinetic operators. Nonetheless, the gauge parameters $\xi^{\mu\nu}_n$ \eqref{gauge_s3} can be used to impose the TT gauge on the rank-3 tensor fields  
\begin{equation}
\label{TT_3}
\nabla_\mu w^{\mu\nu\rho}_n = 0\,,\qquad w^\mu_n = 0\,,\qquad n=0,1,2,..., \infty\,.
\end{equation}
Combining the TT gauge conditions \eqref{TT_3} and the Schouten  identities \eqref{Schouten3} one finds that
\be
\ba{l}
\dps
\Box w^{\mu\nu\rho}_n - \nabla^{(\mu}\nabla_\sigma w^{\nu\rho)\sigma}_n + \frac{1}{2}\nabla^{(\mu}\nabla^{\nu}w^{\rho)}_n - g^{(\mu\nu}\Box w^{\rho)}_n +  g^{(\mu\nu}\nabla_\sigma\nabla_\zeta w^{\rho)\sigma\zeta}_n - \frac{1}{2}g^{(\mu\nu}\nabla^{\rho)}\nabla_\sigma w^\sigma_n 
\\
\\
= 3aw^{\mu\nu\rho}_n\;,\qquad n=0,1,2,..., \infty\,,
\ea
\ee
where the left-hand side is the first three lines in the equation of motion \eqref{e1}  containing  second derivatives of $w^{\mu\nu\rho}_n$. Reorganizing the  equations of motion \eqref{e1}-\eqref{e5} and redefining the scalar field as ${a}^{-1}\sqrt{2/3}\,\phi_0 \to \phi_0$, we finally obtain a simplified  system 
\begin{align}
&\begin{cases}\label{eq:s3eom1}
n \geq 0: \quad \mm[3]{n}\big[-2g^{(\mu\nu}\nabla_\sigma h^{\rho)\sigma}_{n+1}+\nabla^{\mu}h^{\nu\rho)}_{n+1}+\frac{1}{2}g^{(\mu\nu}\nabla^{\rho)}h_{n+1}\big] -\\
\phantom{n \geq 0: \quad}- a((\mm[3]{n})^2-1)w^{\mu\nu\rho}_n = 0\,,\\
n \geq 1: \quad -6h^{\mu\nu}_n + \frac{3}{2}((\mm{n})^2+2)g^{\mu\nu}h_n  = 0\,,
\end{cases}\\
&\;\;\;\;\Box A^\mu_0 - \nabla^\mu\nabla_\nu A^\nu_0 - 5a A^\mu_0 = 0\,,\label{eq:s3eom2}\\
&\begin{cases}\label{eq:s3eom3}
\big[ \nabla^\mu\nabla^\nu - g^{\mu\nu}(\Box - 7a) \big]\phi_0 + 6a\big[ -h^{\mu\nu}_0 + g^{\mu\nu}h_0 \big] = 0\,,\\
\big[ \Box  - 11a \big]\phi_0 - \frac{3}{2}\big[ \nabla_\mu\nabla_\nu - g_{\mu\nu}(\Box - 7a) \big]h^{\mu\nu}_0 = 0\,.
\end{cases}
\end{align}

From the first subsystem  \eqref{eq:s3eom1} it follows that the modes $h^{\mu\nu}_n$ with $n=1,2,..., \infty$ and $w^{\mu\nu\rho}_{n}$ with $n=0,1,2,..., \infty$ vanish on-shell thereby carrying no degrees of freedom. The second subsystem \eqref{eq:s3eom2} is the Proca equation which can be equivalently represented as 
\be
\big[ \Box - 5a \big]A^{\mu}_0 = 0\,,
\qquad
\nabla_\mu A^\mu_0 = 0\,.
\ee 
The third subsystem \eqref{eq:s3eom3} is not diagonal. However, it can be solved  similar  to the spin-2 case. Taking the trace in the  first equation of \eqref{eq:s3eom3} we find 
\begin{equation}
h_0 = \frac{1}{6a}\big[ \Box - 14a \big]\phi_0\,. 
\end{equation}
Then, substituting this expression  back into \eqref{eq:s3eom3} one can express the rank-2 tensor field $h^{\mu\nu}_0$ in terms of the scalar field $\phi_0$ as  
\begin{equation}
h^{\mu\nu}_0 = \frac{1}{6a}\big[ \nabla^\mu\nabla^\nu - 7ag^{\mu\nu} \big]\phi_0\,. 
\end{equation}  
Finally, using this expression in the second equation of \eqref{eq:s3eom3} one finds a forth-order expression 
\begin{equation}
\big[ \Box  - 11a \big]\phi_0 - \frac{1}{4a}\big[ \nabla_\mu\nabla_\nu\nabla^\mu\nabla^\nu - 7a\Box  - (\Box - 7a)(\Box - 14a ) \big]\phi_0 = 0\,,
\end{equation}
where all  fourth order derivatives cancel each other so that the resulting expression is the  standard second-order equation
\begin{equation}
-\frac{9}{4}\big[ \Box - 6a \big]\phi_0 = 0\,.
\end{equation}

To summarize, we obtain the infinite sequences of modes  vanishing on-shell  
\begin{equation}
\ba{c}
\dps
w^{\mu\nu\rho}_{n} = 0\,,\quad n=0,1,2\ldots,\infty\,,
\vspace{2mm}
\\
\dps
h^{\mu\nu}_n = 0\,,\quad A^\mu_n = 0\,, \quad \phi_n = 0\,,\quad  n=1,2,3\ldots,\infty\,,
\ea
\ee
along with the zeroth rank-2 mode which is an auxiliary field
\begin{equation}
h^{\mu\nu}_0 = \frac{1}{6a}\big[ \nabla^\mu\nabla^\nu - 7ag^{\mu\nu} \big]\phi_0\,.
\end{equation}
The remaining fields  are again the  scalar and vector zeroth modes subjected to the second-order equations
\be
\ba{l}
\dps
\big[-\Box + 6a \big]\phi_0 = 0\,,
\vspace{2mm}
\\
\dps
\big[-\Box + 5a \big]A^{\mu}_0 = 0\,,
\quad 
\nabla_\mu A^\mu_0 = 0\,,
\ea
\ee
which are the Klein-Gordon and  Proca equations, respectively.
Fixing $a=1$ we conclude from \eqref{KGMP} that $E=3$ and PDoF are organized into two $\ao12$ Verma modules $\cD_{3}\oplus \cD_{3}$ according to the branching rules \eqref{branch_list} for $s=3$. Cf.  with the equations \eqref{res_1}-\eqref{res_4}.

\section{Conclusion and outlooks}
\label{sec:conc}

We have analyzed the \ldegr dimensional degression and showed that a given spin-$s$ massless theory  in \3ads degresses into the sum of the  Klein-Gordon and Proca theories in \2ads. Thus, contrary to higher-dimensional Kaluza-Klein type theories the spectrum of a bottom  theory in the \ldegr degression is finite. Moreover, a top  theory in this case is topological while a bottom theory propagates local degrees of freedom. This is consistent with the branching rules showing how a respective $\lao22$ representation decomposes into a direct sum of  $\ao12$ Verma modules.

Note that our definition of a topological field theory in $d+1$ dimensions is that there are no propagating (local) degrees of freedom, i.e. the corresponding field equations are solved in terms of arbitrary functions of less than $d$ continuous variables. A useful concept here is the Gelfand-Kirillov dimension $\#_\mathbb{GK}$ \cite{Gelfand1966} that provides an interface between the representation theory and partial differential equations (in the present context see e.g. \cite{Joung:2014qya,Joung:2019wwf}). For AdS$_3$ topological systems the Gelfand-Kirillov dimension of the spin-$s$ (partially-)massless $\lao22$ representations equals $\#_\mathbb{GK}=1$, while $\#_\mathbb{GK}=2$  corresponds to massive representations, cf. the summary list  \eqref{branch_list}. It is instructive to compare with the Jackiw-Teitelboim (JT) gravity in two dimensions where $\#_\mathbb{GK}=0$ which means there are only finitely many constants parameterizing the space of solutions. We conclude that generally a  topological top theory nonetheless encodes non-trivial degrees of freedom to be interpreted as true dynamical degrees of freedom in a bottom  theory. The \ldegr degression is merely  a manifestation of this fact. And, conversely, local degrees of freedom of some dynamical theory can be uplifted to higher dimensions to be described in terms of  a topological higher-dimensional  theory which is presumably simpler and more  tractable.               

Note that the standard formulation in terms of (gauge) tensor  fields discussed in this paper can be equivalently reformulated in terms of $\cG$-connections with Chern-Simons or BF actions for some gauge algebra $\cG$. Moreover, such a formulation is  known to be extremely useful in  building  and analyzing  interactions. It would be important to understand the AdS compactification/degression directly in terms of connections. Some discussion in this direction can be found in \cite{Achucarro:1993fd,Guralnik:2003we,Grumiller:2003ad} (see also recent \cite{Mertens:2018fds,Gaikwad:2018dfc,Narayan:2020pyj}). One of most interesting models here is the AdS$_3$ higher-spin gravity \cite{Blencowe:1988gj,Campoleoni:2010zq,Henneaux:2010xg} and its $\text{sl}(3)\oplus \text{sl}(3)$ version \cite{Campoleoni:2010zq}  that can also be reformulated perturbatively in terms of  spin-2 and spin-3 tensor fields \cite{Campoleoni:2012hp,Fredenhagen:2014oua}. Since the interacting theory here also remains topological, it is expected that its dimensional degression  yields a version of the AdS$_2$ higher-spin gravity \cite{Alkalaev:2013fsa,Grumiller:2013swa,Alkalaev:2014qpa} which describes finitely many massive spinless excitations with higher-spin-gravitational interactions.\footnote{In particular, the equivalence of $2d$ and $3d$ partition functions of higher-spin fields in the near-horizon region of the near-extremal BTZ black hole  was recently shown in \cite{Datta:2021efl}.}

Another implication of the AdS$_3$/AdS$_2$ degression  may be relevant in the context of AdS$_2$ higher-spin gravity with infinitely many massive spinless  excitations  governed by the higher-spin algebra $hs[\lambda]$, where a real parameter $\lambda$ defines an equidistant mass spectrum \cite{Alkalaev:2019xuv,Alkalaev:2020kut}. The \ldegr spectrum suggests that \2ads modes  may arise in the form of scalar/vector covariant fields which are indistinguishable on-shell but may participate in different types of interactions. Also, it becomes relevant in searching  a bulk theory in the context of AdS$_2$/SYK correspondence, where the bulk degrees are given by an infinite tower of massive spinless  modes \cite{Gross:2017aos,Gross:2017hcz}. In particular, \ldegr degressing $\phi^3$ scalar theory yields an infinite tower of \2ads scalars with certain coupling constants parametrized by the triple overlap integrals which differ however from the SYK spectrum. It may imply  that a bulk theory is more complex, presumably a single scalar field should be replaced with an infinite tower of higher-spin massless AdS$_3$ fields. Then,  each of them yields a couple of spinless massive modes in \2ads thereby producing an infinite massive  spectrum as well.     

In this respect,  let us  mention the CFT bootstrap analogy used to control  the high-energy behaviour of bottom theories  with infinitely many fields arising through the Kaluza-Klein reduction of Yang-Mills and Einstein theories in higher dimensions  \cite{Bonifacio:2019ioc,Bonifacio:2020xoc}. This is suggested by the observation that the coupling constants  of a bottom theory are given by multiple overlap integrals subjected to the bootstrap-like  constraints. Our results revealing finite spectra suggest that the bootstrap analogy introduces here the class of “Kaluza-Klein minimal models". Hopefully, it may provide an example of exact solutions to the unitary sum rules  in bottom theories provided they are degressed from a topological top  theory.

To conclude, it would be important to develop the AdS$_3$/AdS$_2$ degression for AdS$_3$ massless rank-$s$  fields $\phi_{m_1...m_s}$ thereby extending the results of the present paper to $s\geq 4$. A non-trivial aspect here would be to take a proper account of  the double-tracelessness condition activated at  $s\geq 4$. Another important  issue is to extend our analysis to  partially-massless spin-$s$ depth-$t$ AdS$_3$ fields relevant in the context of     respective  interacting theories \cite{BUCHBINDER2012243,PhysRevD.102.066003}. Also, it would be interesting to study the degression for  continuous spin AdS$_3$ fields    \cite{Zinoviev:2017rnj}. 

Finally, let us remark that when considering the \ldegr degression of the linearized Einstein gravity with the cosmological term $\Lambda$ in Section  \bref{sec:ads_grav} we did not see any trace of the  linearized JT theory which presumably should arise in $2d$ gravitational systems. We expect that the JT gravity will show up either when considering the AdS$_3$ waveguide compactification with an extra parameter $\alpha$ measuring the size of extra dimension (see \eqref{janus}; in this case partially-massless fields may arise \cite{Gwak:2016sma}), or one should start with the (linearized) AdS$_3$ partially-massless gravity instead. Both options agree with our current understanding that higher-spin fields in \2ads are to be interpreted as partially-massless \cite{Alkalaev:2013fsa}.

\vspace{3mm}

\paragraph{Acknowledgements.} We are grateful to X. Bekaert,  K. Hinterbichler, E. Joung, R. Metsaev, K. Mkrtchyan  for  useful correspondence and discussions. Our work was supported by  the Foundation for the Advancement of Theoretical Physics and Mathematics “BASIS”.

\appendix
\section{Notation and conventions for AdS spaces}
\label{app:notation}

In this Appendix we describe our notation and conventions for AdS spaces. Consider first the AdS$_{d+1}$ spacetime as a hyperboloid embedded in the ambient spacetime $\mathbb{R}^{2,d}$ 
\begin{equation}
-Y^2_{\bar 0}-Y^2_0+Y^2_1+\ldots +Y^2_d = -\frac{1}{b}\;,
\end{equation}
where the parameter $b>0$ will correspond to the constant negative scalar curvature of AdS$_{d+1}$.  
A foliation into AdS$_d$ slices  can be parametrized as follows \cite{Artsukevich:2008vy} 
\begin{equation}
Y_{\bar 0} = \sqrt{\frac{a}{b}}\cosh z\, y_{\bar 0}(x)\;,\quad \ldots\;,\quad Y_{d-1} = \sqrt{\frac{a}{b}}\cosh z\, y_{d-1}(x)\;,\quad Y_d = \frac{1}{\sqrt{b}}\sinh z\;,
\end{equation}
where $a>0$ is a constant associated with the AdS$_d$ scalar curvature, $x=\{ x^\mu, \mu = 0,\ldots d-1\}$ are local coordinates in AdS$_d$ spacetime and $z$ is the slicing coordinate. AdS$_d$ spacetime can also be represented as a hyperboloid in the ambient spacetime $\mathbb{R}^{2,d-1}$ of one less dimension 
\begin{equation}
-y^2_{\bar 0}-y^2_0+y^2_1+\ldots +y^2_{d-1} = -\frac{1}{a}\;.
\end{equation}
This construction yields  the following block-diagonal form of the metric $G_{mn}$ in AdS$_{d+1}$ (indices $m$, $n$ run $0,\ldots,d$)
\begin{equation}
\label{eq:apanz}
G_{mn}(x,z)=
\frac{1}{b}\left(
\begin{array}{c|c}
a\cosh^2 z g_{\mu\nu}(x) & 0 \\
\hline
0 & 1
\end{array}
\right)\;,
\end{equation}
where $g_{\mu\nu}(x)$ is the metric in AdS$_d$ ($\mu \equiv m = 0,\ldots,d-1$). The respective integration measures read
\be
\label{int_meas_1}
d\mu_d = d^dx\,\sqrt{-g}\,,
\qquad
d\mu_{d+1} = d^d x dz\, \sqrt{-G} = d\mu_d dz\, a^{d/2}b^{-(d+1)/2}  \cosh^dz \;,
\ee
where $G$ and $g$ are the determinants of the AdS$_{d+1}$ and AdS$_{d}$ metrics,  respectively. 

The $(d+1)$-th value of an index $m$ is denoted by $\bullet \equiv  m = d$. Note that the metric  \eqref{eq:apanz}  becomes identical to \eqref{av_metric} provided that  $a=b=\ell^{-2}_{_{AdS}}$, i.e. both original AdS$_{d+1}$ and its slices AdS$_d$ have the same radius. The reason behind keeping parameters $a$ and $b$ arbitrary  is to control mass-like terms in the respective actions of  AdS$_{d}$ and AdS$_{d+1}$ theories.  

The Riemann tensor and covariant derivatives are defined as
\begin{equation}
{\ol{R}^m}_{nkl} = \partial_{[k} \ol \Gamma_{l]n}^{m} + \ol \Gamma^m_{p[k}\ol \Gamma^p_{l]n}\;,
\end{equation}
 \begin{equation}
 \left[  \ol \nabla_m, \ol\nabla_n  \right] {T^{k\cdots}}_{l\cdots} = \ol R_{mn\phantom{k}p}^{\phantom{mn}k}{T^{p\cdots}}_{l\cdots} +\ldots
 + \ol R_{mnl}^{\phantom{mnl}p}{T^{k\cdots}}_{p\cdots}+\ldots\;,
 \end{equation}
where the  bars over the Riemann tensor, Christoffel symbols and covariant derivatives refer to AdS$_{d+1}$ geometry, otherwise this is AdS$_d$ geometry. The brackets $(mn\ldots)$ and $[mn\ldots]$ denote (anti)symmetrization of indices, which are defined as a sum of essentially different  terms with a unit weight. Then, the scalar curvatures of AdS$_{d+1}$ and AdS$_d$ along with the Christoffel symbols read
\begin{equation}
\bar R = -bd(d+1)\;,\quad  R = -ad(d-1)\;,
\end{equation}
\begin{equation}
\overline{\Gamma}^\rho_{\mu\nu} = \Gamma^\rho_{\mu\nu}\;,\quad \overline{\Gamma}^\mu_{\nu\bullet} = \tanh z\,\delta^\mu_\nu\;, \quad \overline{\Gamma}^{\bullet}_{\mu\nu} = -a\cosh^2 z\tanh z\,g_{\mu\nu}\;.
\end{equation}
Note that  curvatures $R$ and $\bar R$ of the original space and its slices are generally different. 

By  way of example, AdS$_{d+1}$ covariant derivatives acting on a rank-$3$ symmetric tensor $W^{mnk}$ with respect to  the metric  \eqref{eq:apanz} is given by  
\be
\ba{l}
\dps
\ol{\nabla}_\alpha W^{\mu\nu\bullet} = \dd_\alpha W^{\mu\nu\bullet} + \ol{\Gamma}^{(\mu}_{\alpha p}W^{\nu)p\bullet} + \ol{\Gamma}^\bullet_{\alpha p}W^{\mu\nu p} = 
\vspace{2mm}
\\
\hspace{16mm}= \nabla_\alpha W^{\mu\nu\bullet} + \tanh z\, \delta_\alpha^{(\mu}W^{\nu)\bullet\bullet}-a\cosh^2 z\tanh z\, {W_{\alpha}}^{\mu\nu}\;,
\vspace{2mm}
\\
\dps
\ol{\nabla}_\bullet W^{\mu\nu\rho} = \dd W^{\mu\nu\rho} + \ol{\Gamma}^{(\mu}_{\bullet p}W^{\nu\rho)p} = (\dd + 3\tanh z)W^{\mu\nu\rho}\;,
\ea
\ee
where we introduced the notation  $\dd \equiv \dd/\dd z$. Also, the Laplace-Beltrami operator in AdS$_d$ spacetime is denoted as $\Box = g^{\mu\nu}\nabla_\mu\nabla_\nu$.
 
In order to simplify our notations  we denote  quadratic combinations of  symmetric tensors $X^{m\ldots}$ in AdS$_{d+1}$ as
\begin{equation}
(\ol\nabla_a X^{m\ldots})^2 \equiv  \ol\nabla_a X^{m\ldots} \ol\nabla^a X_{m\ldots}\;,
\qquad  (X^{m\ldots})^2 \equiv  X^{m\ldots} X_{m\ldots}\;.
\end{equation} 
The same notation is used for AdS$_d$ tensors and their derivatives.

\section{Discrete transformations and Lie algebra representations }
\label{app:dec}

\subsection{AdS$_3$ isometry group} 

AdS$_3$ isometry group is given by the split indefinite orthogonal Lie group $\lgo22$ which leaves invariant $\eta_{AB} =\text{diag}(-1,-1,1,1)$, where $A,B = \bar 0,0,1,2$. There are four  components; the identity component $\lgop22$ yields   the  Lie algebra $\lao22$.  The discrete transformations are generated by the elements
\be
\label{PT}
\text{T} = \text{diag}(1,-1,1,1)\;,
\qquad
\text{P} =  \text{diag}(1,1,-1,1)\;,
\ee 
which correspond to time reversal and space reflection (parity transformation). In addition to $\lgop22$ the three other  components are generated from $\lgop22$ by acting on its elements with $\text{T}$, $\text{P}$,  $\text{PT}$. The component $\text{PT}\cdot \lgop22 \subset \text{SO}(2,2)$. The commutation relations of the Lie algebra $\lao22$ read 
\be
\label{commutation}
i\,[M_{AB}, M_{CD}] = \eta_{BC} M_{AD}-\eta_{AC} M_{BD}+\eta_{AD} M_{BC}-\eta_{BD} M_{AC}\;.
\ee 
The generators are all Hermitian $M^\dagger_{AB} = M_{AB}$. 

To build a highest-weight or lowest-weight  (generalized) Verma module the six $\lao22$ generators are rearranged as
\be
\hat E = M_{\bar 0 0}\;, 
\qquad 
\hat S = M_{12}\;,   
\qquad
\hat L_{\pm k} = M_{0k} \pm i M_{\bar0 k}\;, \quad k=1,2\;,  
\ee
which are, respectively,  energy,  spin,  lowering/raising boosts. The generators are conjugated as  $\hat E^\dagger = \hat E$,\; $\hat S^\dagger = \hat S$,\; $(\hat L_{+k})^\dagger = \hat L_{-k}$, while the discrete symmetries \eqref{PT} act as 
\be
\label{T}
\text{T}\, \hat E\, \text{T}^{-1} = \hat E\;,
\qquad
\text{T}\, \hat S \,\text{T}^{-1} = -\hat S\;,
\qquad
\text{T}\, \hat L_{\pm k} \,\text{T}^{-1} = \hat L_{\pm k}\;,
\ee
\be
\label{P}
\qquad\text{P}\, \hat E \,\text{P}^{-1} = \hat E\;,
\qquad
\text{P}\, \hat S\, \text{P}^{-1} = -\hat S\;,
\qquad
\text{P}\, \hat L_{\pm k} \,\text{P}^{-1} = (-)^k \hat L_{\pm k}\;.
\ee
where T and P are,  respectively,  antilinear and linear operators  \cite{wigner1931gruppentheorie}.\footnote{For unitary representations these operators become  (anti)-unitary. On the other hand, assuming that both $\text{T}$ and $\text{P}$ are linear we find that   highest-weight and lowest-weight conditions (i.e. positive/negative discrete series) are interchanged. }  

A {\it lowest-weight} state  $|E,s\rangle$ spans an irreducible representation of $\cao2\oplus \cao2$ subalgebra of $\hat E$ and $\hat S$ (two-dimensional space) 
\be
\label{LW}
\hat E|E,s\rangle = E|E,s\rangle\;, 
\qquad 
\hat S|E,s\rangle = s |E,s\rangle\;,
\qquad
\hat L_{- k}|E,s\rangle = 0\;,
\ee
where the energy $E\in \mathbb{R}$,  the spin is  integer $s\in \mathbb{Z}$ (we consider only bosonic modules).
A lowest-weight  module is then generated by acting with basis monomials of the raising boosts as
\be
\label{genVerma}
\cV_+(E,s) = \{\,\hat L_{+ k_1}\hat L_{+ k_2} \cdots \hat L_{+ k_l}|E,s\rangle,\;\; l=0,1,2,...\,\}\;.
\ee  
Here, $l$ denotes the level and each level-$l$ space is organized into $\cao2$ finite-dimensional irreps of spin numbers in the range $|s+l|, |s+l-2|, ..., |s-l|$. For lowest-weight representations the energy  is bounded from below. At specific values $E = E_0(s,t)$ there are singular vectors arising on the $(t+1)$-th level. The resulting $\lao22$ quotient modules  $\cH(E_0,s)$  are described in Section \bref{sec:branch}. 

Let us now discuss how discrete symmetries \eqref{T}-\eqref{P}  act on the modules $\cV_+(E,s)$ \eqref{genVerma}. The energy operator is invariant under all discrete transformations P, T, PT. On the contrary, the spin operator changes the sign under P and T, and remains PT invariant. One can define the vacuum  $| \widetilde{ E, s}\rangle \coloneqq| E, -s\rangle \oplus |E,s\rangle$ which stays   invariant under all discrete symmetries. Recalling  how the discrete symmetries act on the boosts \eqref{T}-\eqref{P} we conclude that other lowest-weight conditions in \eqref{LW} remain intact. Thus,  a module generated from a discrete-invariant vacuum  is the following direct sum of (generalized) Verma modules  
\be
\label{main_verma}
\cD(E,s) = \cV_-(E,-s)\oplus \cV_+(E,s)\;,
\ee       
with a lowest-weight space containing   states of opposite spins $\pm s$. From now on we set the spin to be non-negative integer $s\in \mathbb{N}$. This definition can be equivalently replaced by imposing a weaker condition $\mathbb{C}_2[\cao2] | \widetilde{ E, s}\rangle = s^2 | \widetilde{ E, s}\rangle$, where $| \widetilde{ E, s}\rangle$ is the lowest-weight state of $\cD(E,s)$ and $\mathbb{C}_2$ denotes the quadratic Casimir operator. It is these modules \eqref{main_verma} which are discussed  in  Section \bref{sec:branch}.

One can consider two other bases in the Lie algebra $\lao22$: the Lorentz basis and the factorized basis.    
The Lorentz basis can be built by the standard block decomposition of an antisymmetric matrix $M_{AB}$ as 
\be
\label{basis1}
P_a = M_{\bar0a}\;, 
\qquad
L_{ab} = M_{ab}\;,
\qquad
a,b = 0,1,2\;, 
\ee
which are momentum and Lorentz rotation generators commuting in the standard fashion. On the other hand, the decomposition of $\lao22$ in simple subalgebras  can be achieved by introducing linear  combinations 
\be
\label{two_copies}
J_a^{(\varepsilon)} = \frac{1}{2}\left(P_a+ \varepsilon M_a\right)\;, 
\quad 
\varepsilon = \pm\;;
\qquad\;
[J_a^{(-)},J_b^{(+)}]=0\;,
\ee
where $M_a = \half \epsilon_{abc}M^{bc}$ and $\epsilon_{012}=1$, cf. \eqref{basis1}. Then, each set of elements $J_a^{(\varepsilon)}$ defines simple factors $\ao12\oplus \text{o}(2,1) = \lao22$.  The discrete symmetries act on them as $\text{P} J_0^{(+)} \text{P}^{-1}= J_0^{(-)}$ and $\text{P} J_0^{(+)} \text{P}^{-1}= J_0^{(-)}$ that means the two factors  are interchanged.  The energy and spin operators are given by 
\be
\label{defES}
\hat E = J_0^{(-)}+J_0^{(+)}\;,
\qquad
\hat S = J_0^{(-)}-J_0^{(+)}\;.
\ee

Each copy of $\ao12$ defines its own  Verma module. To this end one introduces diagonal and  raising/lowering operators as
\be
\label{o21}
J_0^{(\varepsilon)}\coloneqq J_0^{(\varepsilon)}\;,
\qquad
J_\pm^{(\varepsilon)} \coloneqq J_1^{(\varepsilon)}\pm i J_2^{(\varepsilon)}\;,
\ee
which define lowest-weight conditions imposed in a lowest-weight state $|h_\varepsilon\rangle$ as 
\be
J_0^{(\varepsilon)} |h_\varepsilon\rangle = h_\varepsilon|h_\varepsilon\rangle\;,
\qquad
J_-^{(\varepsilon)} |h_\varepsilon\rangle = 0\;,
\ee
for some $h_\varepsilon \in \mathbb{R}$. A lowest-weight $\ao12$ representation is built as
\be
\label{o12verma}
\cD_{h_\varepsilon} = \left\{\, \Big(J_+^{(\varepsilon)}\Big)^m|h_\varepsilon\rangle,\;\; m=0,1,2,3,...\,, \varepsilon  = \pm\right\}\;. 
\ee

Now, lowest-weight $\lao22$ representations can be constructed by tensoring respective $\ao12$ representations \eqref{o12verma}. Introducing  a lowest-weight vector $|h_-\rangle\otimes |h_+\rangle$ one  constructs  $\lao22$ representation $\cD_{h_-}\otimes \cD_{h_+}$. Here, a basis energy and spin are parametrized as $E = h_-+h_+$  and $s = |h_- -h_+|$, cf. \eqref{defES}. The discrete transformations interchange the weights as  $h_- \leftrightarrow h_+$ and so $E\to E$,  $s \to - s$. It follows that P and T invariant representations take a duplicated form $\left(\cD_{h_-}\otimes \cD_{h_+}\right)\oplus \left(\cD_{h_+}\otimes \cD_{h_-}\right)$ which is in fact \eqref{main_verma}.

\subsection{AdS$_2$ isometry group} 

AdS$_2$ isometry group is given by the indefinite orthogonal Lie group $\algo22$ which leaves invariant $\eta_{AB} =\text{diag}(-1,-1,1)$, where $A,B = 0',0,1$ (see e.g. \cite{Bargmann:1946me,Barut1965}). It has four  components; the identity component $\algop22$ yields  the  Lie algebra $\text{o}(2,1)$.  The discrete transformations are generated by the elements
\be
\label{PT2}
\text{T} = \text{diag}(1,-1,1)\;,
\qquad
\text{P} =  \text{diag}(1,1,-1)\;,
\ee 
which correspond to time reversal and space reflection (parity transformation). Here, we kept the same notation as in \eqref{PT}. The commutation relations of the Lie algebra $\text{o}(2,1)$ 
\be
\label{commutation2}
i\,[M_{A}, M_{B}] =  - \epsilon_{ABC} \eta^{CD}M_D\;,
\ee 
can be obtained from \eqref{commutation} when $A,B,C$ are three-dimensional by substitution $M_A = \half \epsilon_{ABC}M^{BC}$, where $\epsilon_{0'01}=+1$.  
The generators are all Hermitian $M^\dagger_{A} = M_{A}$. 

To build a highest-weight or lowest-weight  Verma module the three $\text{o}(2,1)$ generators are rearranged as 
\be
\hat E = M_{0' 0}\;, 
\qquad
\hat L_{\pm} = M_{01} \pm i M_{0' 1}\;,   
\ee
which are  the energy and  raising/lowering boosts (no spin operators in this case). The generators are conjugated as  $\hat E^\dagger = \hat E$ and $(\hat L_{+})^\dagger = \hat L_{-}$, while the discrete symmetries \eqref{PT2} act as 
\be
\ba{c}
\label{PTE}
\text{T}\, \hat E\, \text{T}^{-1} = \hat E\;, 
\qquad\quad 
\text{T}\, \hat L_{\pm}\, \text{T}^{-1} = -\hat L_{\pm}\;,
\vspace{2mm}
\\
\text{P}\, \hat E \,\text{P}^{-1} = \hat E\;, 
\qquad\quad
\text{P}\, \hat L_{\pm} \,\text{P}^{-1} = -\hat L_{\pm}\;,
\ea
\ee
where T and P are, respectively,  antilinear  and linear  operators. One can introduce $\ao12$ Verma lowest-weight modules $\cD_{E}$  \eqref{o12verma} with the lowest energy $E$.  From \eqref{PTE} it follows that $\algo22$ discrete symmetry leaves $\cD_{E}$ invariant so that no duplication similar to \eqref{main_verma} is required.

Consider now the Clebsch-Gordan problem for representations $\cD_E$ and $\nD_E$ (see Section \bref{sec:branch}).  For the coupling of such representations we have \cite{Pukanszky1961,Mukunda:1974gb,repka,zbMATH03701150,vanTonder:2002gh,Sellaroli:2014ega,Basile:2018dzi}
\be
\hspace{-5mm}\cD_{h_1} \otimes \cD_{h_2}\;\;\; =\;\;\; \bigoplus_{h=h_1+h_2}^{\infty}\, \cD_{h}\;,
\ee
\be
\label{DH}
\qquad\;\;\;\;\quad\cD_{h_3} \otimes \cH_{h_4}\;\; =\;\;\;\bigoplus_{h=h_3+h_4}^{h_3-h_4}\, (-)^{h_3+h_4-h}\, \cD_{h}\;,
\ee
\be
\label{HH}
\;\;\;\;\qquad\quad\cH_{h_4} \otimes \cH_{h_5}\;\; =\;\;\; \bigoplus_{h=h_4+h_5}^{-\abs{h_4-h_5}}\, (-)^{h_4+h_5-h}\,\cH_{h}\;,
\ee
for $\forall h_1,h_2$; $h_3>0$, $h_3>\abs{h_4}$, and $h_4,h_5\in \frac{1}{2}\mathbb{Z}_{\hspace{-1mm}\leq\hspace{-0.8mm} 0}$; the summation  goes with a step 1. An essential property here is that $\cD_{h_3} \otimes \cH_{h_4}$ has finitely many components similar to $\cH_{h_4} \otimes \cH_{h_5}$. In the last two products \eqref{DH} and \eqref{HH} we added the sign factors which mean that two modules are summed as $V_a\oplus (-V_b)$, where $-V_b$ has the opposite signature inner product \cite{vanTonder:2002gh}: the flipping  sings in the tensor products with  finite-dimensional modules indicate inevitable   non-unitarity except for the identity module $\cH_0$ of zeroth weight. In other cases, for simplicity,  we ignore the  sign-flipping, if any.

\section{Jacobi polynomials and basis functions} 
\label{app:jacobi}

Here, we recap some basics about the Jacobi polynomials, see e.g. \cite{Shen2011}. Also, we explicitly compute some of the  integral overlaps arising in the dimensional degression. 

The Jacobi polynomials  $J^{\alpha,\beta}_n(x)$ ($n\in \mathbb{N}_0$) are  polynomials in the domain $x\in (-1,1)$ with two real parameters $\alpha,\beta > -1$. They are orthogonal with respect to the Jacobi weight function $\omega^{\alpha,\beta} = (1-x)^\alpha(1+x)^\beta$, namely
\begin{equation}
\label{C1}
\int_{-1}^{1} J_{n}^{\alpha, \beta}(x) J_{m}^{\alpha, \beta}(x) \omega^{\alpha, \beta}(x) d x=\left\|J_{n}^{\alpha, \beta}\right\|^2 \delta_{m n}\,,
\end{equation}
where 
\begin{equation}
\left\|J_{n}^{\alpha, \beta}\right\|^2=\frac{2^{\alpha+\beta+1} \Gamma(n+\alpha+1) \Gamma(n+\beta+1)}{(2 n+\alpha+\beta+1) n ! \Gamma(n+\alpha+\beta+1)}\,.
\end{equation}
The Jacobi polynomials are the eigenfunctions of the following  Sturm-Liouville operator 
\begin{equation}
\ba{c}
\mathscr{L}_{\alpha, \beta} J^{\alpha,\beta}_n(x) \equiv -(1-x)^{-\alpha}(1+x)^{-\beta} \partial_{x}\left[(1-x)^{\alpha+1}(1+x)^{\beta+1} \partial_{x} J^{\alpha,\beta}_n(x)\right]
\vspace{2mm}
\\
=\left(x^{2}-1\right) \partial_{x}^{2} J^{\alpha,\beta}_n(x)+\big(\alpha-\beta+(\alpha+\beta+2) x\big) \partial_{x} J^{\alpha,\beta}_n(x) =\lambda_{n}^{\alpha, \beta} J_{n}^{\alpha, \beta}(x)\,,
\ea
\end{equation}
where the eigenvalues are $\lambda_{n}^{\alpha, \beta}=n(n+\alpha+\beta+1)$. The Jacobi polynomials are given by 
\begin{al}
J^{\alpha,\beta}_n(x) &= \frac{\Gamma(n+\alpha+1)}{n ! \Gamma(n+\alpha+\beta+1)} \sum_{k=0}^{n}\binom nk \frac{\Gamma(n+k+\alpha+\beta+1)}{\Gamma(k+\alpha+1)}\left(\frac{x-1}{2}\right)^{k} =\\
&= \frac{\Gamma(n+\alpha+1)}{n ! \Gamma(\alpha+1)} {}_2F_1(-n,n+\alpha+\beta+1;\alpha+1;\frac{1-x}{2})\,.
\end{al}
The Jacobi polynomials can be defined by the recurrence equations 
\begin{al}\label{eq:aprecJ1}
&J^{\alpha,\beta}_0(x) = 1\,, \quad J^{\alpha,\beta}_1(x) = \frac{1}{2}(\alpha+\beta+2)x + \frac{1}{2}(\alpha-\beta)\,,\\
&J^{\alpha,\beta}_{n+1}(x) = (a_n x-b_n)J^{\alpha,\beta}_{n}(x) - c_nJ^{\alpha,\beta}_{n-1}(x)\,,\quad n\geq 1\,,
\end{al}
where
\begin{equation}
\label{eq:aprecJ2}
\begin{aligned}
&a_{n}=\frac{(2 n+\alpha+\beta+1)(2 n+\alpha+\beta+2)}{2(n+1)(n+\alpha+\beta+1)}\,, \\
&b_{n} = \frac{\left(\beta^{2}-\alpha^{2}\right)(2 n+\alpha+\beta+1)}{2(n+1)(n+\alpha+\beta+1)(2 n+\alpha+\beta)}\,, \\
&c_{n} =\frac{(n+\alpha)(n+\beta)(2 n+\alpha+\beta+2)}{(n+1)(n+\alpha+\beta+1)(2 n+\alpha+\beta)}\,.
\end{aligned}
\end{equation}
Note that in this paper we use the Jacobi polynomials with $\alpha = \beta$ so that  $b_n = 0$ for any $n$. The following  differential constraint satisfied by the Jacobi polynomials is useful in practice
\begin{equation}
\label{eq:apdJ}
\partial_x J^{\alpha,\beta}_{n}(x) = \mu^{\alpha,\beta}_{n}J^{\alpha+1,\beta+1}_{n-1}(x)\,,\quad \mu^{\alpha,\beta}_{n}=\frac{1}{2}(n+\alpha+\beta+1)\,.
\end{equation}

Two families  of the Jacobi polynomials with parameters $\alpha,\beta$ and $a,b$ can be related to each other by the linear transformation 
\begin{equation}
\label{eq:apJJ1}
J_{n}^{\alpha, \beta}(x)=\sum_{k=0}^{n} \hat{c}_{k}^{n} J_{k}^{a, b}(x)\,,
\end{equation}
where the transition coefficients are given by 
\begin{equation}
\label{eq:apJJ2}
\begin{aligned}
\hat{c}_{k}^{n}=& \frac{\Gamma(n+\alpha+1)}{\Gamma(n+\alpha+\beta+1)} \frac{(2 k+a+b+1) \Gamma(k+a+b+1)}{\Gamma(k+a+1)}\times \\
& \times \sum_{m=0}^{n-k} \frac{(-1)^{m} \Gamma(n+k+m+\alpha+\beta+1) \Gamma(m+k+a+1)}{m !(n-k-m) ! \Gamma(k+m+\alpha+1) \Gamma(m+2 k+a+b+2)}\,.
\end{aligned}
\end{equation}

\subsection{Basis functions}
\label{app:basis}
To perform  the \ldegr degression we use the following basis functions  built in terms of the Jacobi polynomials  
\be
\ba{l}
\dps
P^s_n(z) = \tilde{N}^s_n(\cosh z)^{-d-2s+2} J^{\frac{d+2s-4}{2},\frac{d+2s-4}{2}}_n(-\tanh z) =
\vspace{2mm}
\\
\dps
\hspace{10.7mm}= N^s_n(\cosh z)^{-d-2s+2} {}_2F_1(-n,n+d+2s-3;\frac{d+2s-2}{2};\frac{1+\tanh z}{2})\,,
\ea
\ee
where the normalization constants read
\be
N^s_n = \tilde{N}^s_n\frac{\Gamma(n+\frac{d+2s-2}{2})}{n!\Gamma(\frac{d+2s-2}{2})} = \frac{\sqrt{d+2s-3+2n}}{2^{\frac{d+2s-3}{2}}\Gamma(\frac{d+2s-2}{2})}\sqrt{\frac{\Gamma(d+2s-3+n)}{n!}}\,.
\ee
Such a set of the basis functions turns out to be convenient when describing the higher-dimensional \hdegr degression \cite{Artsukevich:2008vy}. 
In the $d=2$ case the parameter  $s$ must be $s>1$ otherwise $\frac{d+2s-4}{2} \leq -1$. It follows that in two dimensions we cannot consider functions $P^0_n$ as  their inner products built using \eqref{C1} do not converge. This is another critical reason why the  two-dimensional case is different from the higher-dimensional case. 

For the later convenience, let us denote 
\begin{equation}
J^s_n \equiv J^{\frac{d+2s-4}{2},\frac{d+2s-4}{2}}_n\,.
\end{equation}
The  functions $\{P^s_n\}_{n\in \mathbb{N}_0}$ defined above form an orthonormal basis
\be
\label{eq:PPdelta}
\ba{c}
\dps
(P^s_n,P^s_m)_s \equiv \int^{+\infty}_{-\infty}dz(\cosh z)^{d+2s-2} P^s_n(z) P^s_m(z) = 
\vspace{2mm}
\\
\dps
= \tilde{N}^s_n\tilde{N}^s_m \int^{+\infty}_{-\infty}dz(\cosh z)^{-d-2s+2} J^{s}_n(-\tanh z) J^{s}_m(-\tanh z) =
\vspace{2mm}
\\
\dps
= \tilde{N}^s_n\tilde{N}^s_m \int^{+1}_{-1}dx(1-x)^{\frac{d+2s-4}{2}}(1+x)^{\frac{d+2s-4}{2}} J^{s}_n(x)J^{s}_m(x) =\delta_{nm}\,.
\ea
\ee
The relation \eqref{eq:PPdelta}  introduces an inner product $(A,B)_s$ between two basis functions. It has the following  obvious properties
\be
(A,B)_s = (B,A)_s\,,
\qquad
(A,\tanh^n zB)_s = (\tanh^n zA,B)_s\,.
\ee

A convenient way to organize terms arising in a top action is to introduce a first-order differential operator $\LL_n$ originating  from AdS$_{d+1}$ covariant derivatives (similar to the one used in \cite{Gwak:2016sma}: $\cos^{-1}\theta$ changed to $\cosh z$)
\begin{equation}
\label{eq:apL}
\LL_nA \equiv \cosh^{-n} z \partial (\cosh^n z A) \equiv (\partial+n\tanh z)A\,.
\end{equation}
The $\LL$-operator has the property that
\be
\label{eq:ap(L)}
(A,\LL_nB)_s = -(\LL_{d+2s-2-n}A,B)_s\,,
\qquad
(A,\cosh^2z\LL_2B)_s = -(\LL_{d+2s-2}A,B)_{s+1}\,.
\ee

The functions $\{P^s_n(z)\}_{n\in \mathbb{N}_0}$ can be defined as  eigenfunctions of the corresponding Sturm-Liouville operator\footnote{In  refs. \cite{Metsaev:2000qb,Artsukevich:2008vy,Gross:2017aos} it was chosen  to be the Pöschl-Teller Hamiltonian.}
\begin{al}
\LL_{d+2 s-4}\left(\cosh ^{2} z \LL_{2} P_{n}^{s}\right)= \big[\cosh ^{2} z \LL_{2}\LL_{d+2 s-2} -(d+2 s-4) \big]P_{n}^{s}  =-\left(\mm[s]{n}\right)^{2} P_{n}^{s}\,,
\end{al}
where
\begin{equation}
\label{eq:m}
\mm[s]{n}=\sqrt{(n+1)(n+d+2 s-4)}\,.
\end{equation}
Thus, these functions form a basis in the $(\cosh z)^{d+2s-2}$-weighted $L^2(\mathbb{R})$ space. 

Using the  equations \eqref{eq:apdJ} and \eqref{eq:ap(L)} one can obtain differential relations between neighbouring families  of the basis functions (parameterized by $s$)
\begin{equation}
\label{LPmP}
\cosh ^{2} z \,\LL_{2} P_{n}^{s}=\mm[s]{n} P_{n+1}^{s-1}, 
\qquad 
\LL_{d+2 s-2} P_{n}^{s}=-\mm[s+1]{n-1} P_{n-1}^{s+1}\,.
\end{equation}
It is worth noting that   $\cosh ^{2} z\, \LL_{2}$ and $\LL_{d+2 s-2}$ can be considered as lowering/raising  operators in $s$ or $n$, respectively.  

Now, using the recurrence relations \eqref{eq:aprecJ1} for the Jacobi polynomials  we obtain  the similar recurrence formula (at $\alpha = \beta = (d+2s-4)/2$)
\begin{equation}
\label{eq:aprecP}
\tanh z\, P^s_n = -\frac{1}{a_n}\frac{\tilde{N}^s_n}{\tilde{N}^s_{n+1}}P^s_{n+1} - \frac{c_n}{a_n}\frac{\tilde{N}^s_n}{\tilde{N}^s_{n-1}}P^s_{n-1}\,,
\end{equation}
where the second term is non-vanishing only if $n\geq 1$.

\subsection{Inner products}\label{app:inn}
The recurrence relation  \eqref{eq:aprecP} allows computing  non-trivial inner products with $\tanh z$ and $\tanh^2 z$ (in general, with $\tanh^m z$ for higher $m$)
\begin{al}
(P^s_n,\tanh zP^s_m)_s &= \begin{cases}
-\frac{1}{a_n}\frac{\tilde{N}^s_{n-1}}{\tilde{N}^s_{n}}\,,\quad &n=m+1\,,\\
-\frac{c_{n+1}}{a_{n+1}}\frac{\tilde{N}^s_{n+1}}{\tilde{N}^s_{n}}\,,\quad &n=m-1\,,m\geq 1\,,
\end{cases}\\
&= \begin{cases}
-\sqrt{\frac{n(n+d+2s-4)}{(2n+d+2s-3)(2n+d+2s-5)}}\,,\quad n=m+1\,,\\
-\sqrt{\frac{(n+1)(n+d+2s-3)}{(2n+d+2s-1)(2n+d+2s-3)}}\,,\quad n=m-1\,,m\geq 1\,.
\end{cases}
\end{al}
\begin{al}
(P^s_n,\tanh^2 zP^s_m)_s &=\begin{cases}
\frac{1}{a_{n-2}a_{n-1}}\frac{\tilde{N}^s_{n-2}}{\tilde{N}^s_{n}}\,,\quad &n=m+2\,,\\
\frac{c_{n+1}}{a_{n}a_{n+1}}+\frac{c_{n}}{a_{n-1}a_{n}}\,,\quad &n=m\,,m\geq 1\,,\\
\frac{c_{1}}{a_{0}a_{1}}\,,\quad &n=m=0\,,\\
\frac{c_{n+1}c_{n+2}}{a_{n+1}a_{n+2}}\frac{\tilde{N}^s_{n+2}}{\tilde{N}^s_{n}}\,,\quad &n=m-2\,,m\geq 2\,,
\end{cases}\\
&= \begin{cases}
\frac{\sqrt{n(n-1)(n+d+2s-4)(n+d+2s-5)}}{(2n+d+2s-5)\sqrt{(2n+d+2s-3)(2n+d+2s-7)}}\,,\quad &n=m+2\,,\\
\frac{2n(n+d+2s-3)+d+2s-5}{(2n+d+2s-1)(2n+d+2s-5)} \,,\quad &n=m\,,\\
\frac{\sqrt{(n+2)(n+1)(n+d+2s-2)(n+d+2s-3)}}{(2n+d+2s-1)\sqrt{(2n+d+2s+1)(2n+d+2s-3)}}\,,\quad &n=m-2\,,m\geq 2\,.\\
\end{cases}
\end{al}
Using \eqref{eq:apJJ1} one can derive ($\alpha=\beta=(d+2s-4)/2$ and $a=b=(d+2s-2)/2$)
\begin{al}
(P^{s+1}_n,P^s_m)_s &= (P^{s+1}_n,\tilde{N}^s_m\sum^m_{k=0}\hat{c}^m_k\frac{1}{\tilde{N}^{s+1}_k}P^{s+1}_k)_{s+1}=\\
&=\hat{c}^m_n\frac{\tilde{N}^s_m}{\tilde{N}^{s+1}_n}\,,\quad n=\lbrace m, m-2 \rbrace\\
&=\begin{cases}
\sqrt{\frac{(n+d+2s-2)(n+d+2s-3)}{(2n+d+2s-1)(2n+d+2s-3)}}\,,&n=m\,,\\
-\sqrt{\frac{(n+2)(n+1)}{(2n+d+2s+1)(2n+d+2s-1)}}\,,&n=m-2\,,m\geq 2\,.
\end{cases}
\end{al}
It turns out that $\hat{c}^m_k$ are non-vanishing only  when $n=m$ and $n=m-2\,,m\geq 2$.

All other relevant  inner products can be obtained  in the similar way. Below we list those ones  used in this paper. Recall that the coefficients $\hat{c}^m_n$ here implicitly depend on $\alpha$, $\beta$, $a$, $b$, cf. \eqref{eq:apJJ2}.  
\be
\ba{c}
\dps
(P^{s}_n,\LL_iP^s_m)_s = (P^{s}_n,[\LL_{d+2s-2}+(i-d-2s+2))\tanh z]P^s_m)_s = 
\vspace{2mm}
\\
\dps
=-\mm[s+1]{m-1}(P^{s}_n,P^{s+1}_{m-1})_s +(i-d-2s+2)(P^{s}_n,\tanh zP^s_m)_s\,.
\ea
\ee
\be
(P^{s+2}_n,P^s_m)_s = (P^{s+2}_n,\tilde{N}^s_m\sum^m_{k=0}\hat{c}^m_k\frac{1}{\tilde{N}^{s+1}_k}P^{s+1}_k)_{s+1}=\hat{c}^m_n\frac{\tilde{N}^s_m}{\tilde{N}^{s+2}_n}\,,\quad n=\lbrace m, m-2,m-4 \rbrace\,.
\ee 
\begin{al}
(P^{s+1}_n,\tanh zP^s_m)_s = (P^{s+1}_n,\tanh z\tilde{N}^s_m\sum^m_{k=0}\hat{c}^m_k\frac{1}{\tilde{N}^{s+1}_k}P^{s+1}_k)_{s+1}\,.
\end{al}
\begin{al}
(P^{s+1}_n&,\cosh^2 z  \LL_i\LL_jP^s_m)_s = \\
&= (\tilde{N}^{s+1}_n\sum^n_{k=0}\hat{c}^n_k \frac{1}{\tilde{N}^s_k} P^{s}_k,\LL_i\LL_jP^s_m)_{s}\,,k=\left\lbrace n, n-2,n-4,...,n-2\left\lfloor\frac{n}{2}\right\rfloor \right\rbrace\,.
\end{al}%
\begin{al}
(P^{s}_n&,\LL_i\LL_jP^s_m)_s =\\
= &(P^{s}_n,[\LL_{i+j-d-2s+2}\LL_{d+2s-2}+(i-d-2s+1)(j-d-2s+2)\tanh^2 z +\\
&+ (j-d-2s+2)]P^s_m)_s = \\
=&-\mm[s+1]{m-1}(P^{s}_n,[\LL_{d+2s}+(i+j-2d-4s+2)\tanh z]P^{s+1}_{m-1})_s +\\
&+(i-d-2s+1)(j-d-2s+2)(P^{s}_n,\tanh^2 zP^s_m)_s + (j-d-2s+2)\delta_{nm}=\\
=&\mm[s+1]{m-1}\mm[s+2]{m-2}(P^s_n,P^{s+2}_{m-2})_s - (i+j-2d-4s+2)\mm[s+1]{m-1}(P^{s}_n,\tanh zP^{s+1}_{m-1})_s+\\
&+(i-d-2s+1)(j-d-2s+2)(P^{s}_n,\tanh^2 zP^s_m)_s + (j-d-2s+2)\delta_{nm}\,.
\end{al}

\section{Schouten identities }
\label{app:rel}

In two dimensions  there are useful identities that follow from the fact that any tensor with three  antisymmetrized indices is identically zero. In Refs. \cite{Conde:2016izb,Mkrtchyan:2017ixk,Kessel:2018ugi}  such identities with over-antisymmetrization were generally  called the  Schouten identities. In particular, antisymmetrizing a product of three Kronecker deltas $\delta^{\mu\nu\rho}_{\alpha\beta\gamma} = \delta^{[\mu}_\alpha\,\delta^\nu_\beta\,\delta^{\rho ]}_\gamma$ and then contracting  with a second derivative of a symmetric rank-2 tensor field $X^{\mu\nu}$ one obtains   
\begin{equation}
\delta^{\mu\nu\rho}_{\alpha\beta\gamma}\,\nabla_\nu\nabla^\beta {X_\rho}^\gamma \equiv 0\;,
\end{equation}
where the first factor vanishes  identically because all indices take  two values,  $\alpha, \beta, \gamma, \mu, \nu, \rho = 0,1$. Raising the index $\alpha$ yields the following relation 
\begin{equation}
\label{schouten}
\Box X^{\mu\nu} - \nabla^{(\mu}\nabla_\rho X^{\nu)\rho} + \nabla^\mu\nabla^\nu X - g^{\mu\nu}\Box X + g^{\mu\nu}\nabla_\rho\nabla_\sigma X^{\rho\sigma} + a\big[ 2X^{\mu\nu} - g^{\mu\nu}X \big] \equiv 0\,,
\end{equation}   
where the algebraic terms arise  from commutating covariant derivatives (see Appendix \bref{app:notation}) and $X = g^{\mu\nu}X_{\mu\nu}$ is the trace. Remarkably,  the left-hand side of \eqref{schouten} represents the linearized Einstein equation in AdS$_2$ with the curvature $R = -2a$. This is yet another manifestation that the Einstein-Hilbert action in two dimensions is topological. 

The analogous Schouten identities exist  for tensor fields of  rank $s = 3,4,...$, while there are no any non-trivial (i.e. containing $\Box X^\mu$) identities  for $s=1$ tensor fields.  For instance, the Schouten identity for a symmetric rank-3 tensor field $X^{\gamma\rho\sigma}$,  
\begin{equation}\label{eq:sch3}
\delta^{\mu\nu\rho}_{\alpha\beta\gamma}\,\nabla_\nu\nabla^\beta {X_\rho}^{\gamma\sigma} \equiv  0\,,
\end{equation} 
is expanded as follows 
\be
\label{Schouten3}
\ba{l}
\dps
6\Box X^{\mu\nu\rho} - 4\nabla^{(\mu}\nabla_\sigma X^{\nu\rho)\sigma} + \nabla^{(\mu}\nabla^\nu X^{\rho)} - 2g^{(\mu\nu}\Box X^{\rho)}  
\vspace{2mm}
\\
\dps
\hspace{15mm}+ 2g^{(\mu\nu}\nabla_\sigma\nabla_\zeta X^{\rho)\sigma\zeta} + a\big[ -18X^{\mu\nu\rho} + 4g^{(\mu\nu}X^{\rho)} \big] \equiv 0\,.
\ea
\ee

Formally, one can say that in two dimensions there are two series of second-order kinetic operators for rank-$s$ tensor fields $X^{\mu_1 ... \mu_s}$: the Schouten series and the Fronsdal series. They are different in general, but coincide at $s=2$ giving rise to the Einstein tensor. Both of them are gauge invariant with respect to $X^{\mu_1 ... \mu_s} \to X^{\mu_1...\mu_s} + \nabla^{(\mu_s}Y^{\mu_1...\mu_{s-1})}$. However, the Schouten identities are trivially invariant since the gauge variation is another Schouten identity with three derivatives acting on $Y^{\mu_1...\mu_{s-1}}$. On the other hand, the Schouten identities can be used to express the Laplace-Beltrami  operator $\Box X^{\mu_1 ... \mu_s}$ in terms of divergences $\nabla_{\nu} X^{\nu\mu_1 ... \mu_{s-1}}$ and algebraic terms $X^{\mu_1 ... \mu_{s}}$ and $g_{\nu\rho}X^{\nu\rho\mu_1 ... \mu_{s-2}}$ that makes the Fronsdal operators to be proportional to $X^{\mu_1 ... \mu_s}$ provided that the TT gauge is imposed.

Finally, the Schouten type identities also exist in higher dimensions $n\geq 3$ when one acts with an antisymmetrized product of $n+1$ Kronecker symbols $\delta^{\mu_1\ldots\, \mu_{n+1}}_{\alpha_1\ldots \,\alpha_{n+1}}$ on a higher derivative combination of a tensor field, e.g.  
\be
\label{schouten_dim}
\delta^{\mu_1\ldots\, \mu_{n+1}}_{\alpha_1\ldots \,\alpha_{n+1}}\,{{X^{\nu_1\ldots\nu_r}}_{\mu_1}}^{\alpha_1} \nabla_{\mu_2}\nabla^{\alpha_2}\ldots\nabla_{\mu_n}\nabla^{\alpha_n}{{X_{\mu_{n+1}}}^{\alpha_{n+1}}}_{\nu_1\ldots\nu_r} \equiv  0\,,
\ee 
where $X^{\mu_1 ... \mu_{r+2}}$ is a totally-symmetric rank-$(r+2)$ tensor (mixed-symmetry tensor  could be also considered).  Performing index contractions the identity \eqref{schouten_dim} can be cast into the form $X^{\mu_1 ... \mu_{r+2}} \Box^{n-1} X_{\mu_1 ... \mu_{r+2}}+ ... \equiv 0$.
Such identities could be useful in studying higher-order theories in higher-dimensional spaces with typical kinetic terms of the form $X \Box^k X+ ...\;$ with some critical relation between  $n$ and   $k$ that makes such theories topological. In this paper these numbers are $n=2$ and $k=1$.

\section{Mode expansion in the  spin-3 theory}
\label{app:spin3}

Having the decomposition  $\Phi^{mnk}(x,z) = \big\{\Phi^{\mu\nu\rho}(x,z),\, \Phi^{\mu\nu}(x,z),\,\Phi^{\mu}(x,z),\, \Phi(x,z)\big\}$ we  introduce new notations for the component fields along with some convenient redefinitions
\be
\label{dec3}
w^{\mu\nu\rho} \coloneqq \Phi^{\mu\nu\rho} + (ad)^{-1}\sech^2 z\, g^{(\mu\nu}\Phi^{\rho)}\;, 
\quad
h^{\mu\nu} \coloneqq \Phi^{\mu\nu}\;, 
\quad
A^{\mu} \coloneqq \Phi^{\mu}\;, 
\quad
\phi \coloneqq \Phi\;.
\ee
The component fields are all traceful with respect to the AdS$_d$ metric. Similarly, one decomposes the gauge parameter $\Xi^{mn}(x,z) =\big\{\Xi^{\mu\nu}(x,z),\, \Xi^{\mu}(x,z),\, \Xi(x,z),\, \Xi(x,z)\big\} $ and redefines the components as 
\be
\xi^{\mu\nu} \coloneqq a^{-1}b\sech^2z \left[ \Xi^{\mu\nu}+(ad)^{-1}\sech^2z\,g^{\mu\nu}\,\Xi\right]\,,
\ee
\be
\xi^\mu \coloneqq a^{-1}b\sech^2z\, \Xi^{\mu} \,, 
\quad
\xi \coloneqq a^{-1}b\sech^2z\,\Xi\,,
\ee
where the gauge parameter $\xi^{\mu\nu}$ is made traceless. Then, the total action decomposed according to \eqref{dec3}
\begin{equation}
\label{tot3}
S = \sum_{m\geq n}\,\sum_{n=0,1,2,3}S_{mn}\,,
\end{equation} 
is built from the following component actions
\begin{align}
S_{33} =& a^2\iint \cosh^4z \Big[ -(\nabla_\alpha w^{\mu\nu\rho})^2+3(\nabla_\mu w^{\mu\nu\rho})^2 - 6\nabla_\mu w_\nu\nabla_\rho w^{\mu\nu\rho} + 3(\nabla_\alpha w^\mu)^2 +\\
& + \frac{3}{2}(\nabla_\mu w^\mu)^2 + a\big[ -(d-3)(w^{\mu\nu\rho})^2 + 6(d-1)(w^\mu)^2  +\notag\\
&+ w_{\mu\nu\rho}\LL_{d+2}(\cosh^2 z \LL_2w^{\mu\nu\rho}) -3w_\mu \LL_{d+2}(\cosh^2z \LL_2w^\mu)  \big]\Big]\;, \notag\\
S_{32} =& 3a^2\iint \cosh^4z \Big[ -4\nabla_\mu w_\nu \LL_{d+2}h^{\mu\nu} + 2\nabla_\mu w^{\mu\nu\rho} \LL_{d+2}h_{\nu\rho}  + \nabla_\mu w^\mu  \LL_{d+2}h \Big] \;,\\
S_{31} =& 6a^2(d+1)d^{-1}\iint \cosh^4z  w_\mu \LL_{d+2}\LL_{d}A^\mu\;,\\
S_{30} =&  3a\int\int \cosh^2z\nabla_\mu w^\mu \LL_{d}\phi   \;,\\
S_{22} =&  3a\iint \cosh^2z \Big[ -(\nabla_\alpha h^{\mu\nu})^2 + 2(\nabla_\mu h^{\mu\nu})^2 -2\nabla_\mu h\nabla_\nu h^{\mu\nu} + (\nabla_\alpha h)^2 +\notag
\\
&+a\Big[ -2d(h^{\mu\nu})^2 +\frac{1}{2}(3d-2)h^2 -\frac{3}{2}h\LL_{d}(\cosh^2z \LL_{2}h)\Big] \Big]  \;,\notag\\
S_{21} =& {12(d+1)a}{d^{-1}}\int\int\cosh^2z \Big[ \nabla_\mu h^{\mu\nu}\LL_{d} A_\nu - \nabla_\mu h\LL_{d} A^\mu  \Big]\;,\\
S_{20} =&  3\iint  2\phi\big[ \nabla_\mu\nabla_\nu h^{\mu\nu} - \Box h\big] +\\
&+ ah  \big[-\cosh^2z \LL_{-2d-2}\LL_d+2(3d+1)\big]      \phi  \;,\notag\\
S_{11} =& -{3(d+1)}{d^{-2}}\iint d\big[ (\nabla_\alpha A^\mu)^2 - (\nabla_\mu A^\mu)^2 \big]  +\\
& + a\Big[ (3d^2+d-4)(A^\mu)^2 + (d+2)A_\mu \LL_{d-2}(\cosh^2z \LL_{2}A^\mu) \Big]	\;,\notag \\ 
S_{10} =& -{6(d+1)}{d^{-1}}\iint\nabla_\mu A^\mu \LL_{-d}\phi\;,\\
S_{00} =& {a^{-1}}\iint 2\cosh^{-2}z (\nabla_\alpha\phi)^2 +\\
&+ \frac{a}{2}\phi\big[   -\LL_{-6d}\LL_d + d(3d-1)\tanh^2z+11d+4      \big]\phi \;,\notag
\end{align}
where $w^\mu \equiv w^{\mu\nu\rho}g_{\nu\rho}$ and $h\equiv h^{\mu\nu}g_{\mu\nu}$. Here, the integration measure is defined as
$$
\label{measure3}
\iint= b^{-2}\int d\mu_{d+1} = {a}^{d/2}{b}^{-(d+5)/2}\int d\mu_d\int dz \cosh^dz\;,
$$
which  differs   from \eqref{measure2} by an additional factor of $b^{-1}$. The gauge transformations \eqref{gauge33}  given in terms of the component fields read
\be
\ba{l}
\label{gauge3}
\delta w^{\mu\nu\rho} = \nabla^{(\mu}\xi^{\nu\rho)} + {2}{d^{-1}}\,\LL_{d+2}\,g^{(\mu\nu}\xi^{\rho)}\,,\\
\delta h^{\mu\nu} = \nabla^{(\mu}\xi^{\nu)} + a\cosh^2z \,\LL_2\,\xi^{\mu\nu} - {d^{-1}}g^{\mu\nu}\,\LL_{-2d}\,\xi\,,\\
\delta A^\mu = \nabla^\mu\xi + 2a\cosh^2z\, \LL_2\,\xi^\mu\,,\\
\delta\phi = 3a\cosh^2z \,\LL_2\,\xi\,.
\ea
\ee

The mode expansions for the component fields and parameters  are chosen to be  
\be
\label{mode_f3}
\ba{c}
\dps
w^{\mu\nu\rho}(x,z) = \sum^\infty_{n=0} w^{\mu\nu\rho}_n(x)P^3_n(z)\,, 
\qquad
h^{\mu\nu}(x,z)= \sum^\infty_{n=0} h^{\mu\nu}_n(x)P^2_n(z)\,,
\vspace{1mm}
\\
\dps
A^\mu(x,z) = \sum^\infty_{n=0} A^{\mu}_n(x)P^1_n(z)\,, 
\qquad
\phi(x,z) = \sum^\infty_{n=0} \phi_n(x)P^1_n(z)\,, 
\ea
\ee
\be
\label{mode_g3}
\xi^{\mu\nu}(x,z) = \sum^\infty_{n=0}\xi^{\mu\nu}_n(x)P^3_n(z)\,,
\quad
\xi^\mu(x,z) = \sum^\infty_{n=0}\xi^\mu_n(x)P^2_n(z)\,,
\quad
\xi(x,z) = \sum^\infty_{n=0}\xi_n(x)P^2_n(z)\,.
\ee
Note that the fields $A^\mu$ and $\phi$ as well as the gauge parameters $\xi^\mu$ and $\xi$ are  expanded with respect to the same basis functions. Plugging  \eqref{mode_f3}-\eqref{mode_g3} into the gauge transformations  \eqref{gauge3} one finds  the  mode transformations 
\be
\ba{l}
\dps \delta w^{\mu\nu\rho}_n = \nabla^{(\mu}\xi^{\nu\rho)}_n - 2d^{-1}\,\mm[3]{n}\,g^{(\mu\nu}\xi^{\rho)}_{n+1}\,,\\
\dps \delta h^{\mu\nu}_n = \nabla^{(\mu}\xi^{\nu)}_n + a\,\mm[3]{n-1}\,\xi^{\mu\nu}_{n-1} - d^{-1}g^{\mu\nu}\,\sum^\infty_{m=0}(P^2_n,\LL_{-2d}P^2_m)_2\,\xi_m\,,\\
\dps \delta A^\mu_n = \nabla^\mu\sum^\infty_{m=0}(P^1_n,P^2_m)_1\,\xi_m + 2a\,\mm{n-1}\,\xi^\mu_{n-1}\,,\\
\delta\phi_n = 3a\,\mm{n-1}\,\xi_{n-1}\,.
\ea
\ee
Then, integrating out the slicing coordinate $z$ yields the total action \eqref{tot3} in the form
\begin{align}
S_{33} =& a^2\int\sum^\infty_{n=0}\Big\{ -(\nabla_\alpha w_n^{\mu\nu\rho})^2+3(\nabla_\mu w_n^{\mu\nu\rho})^2 - 6\nabla_\mu w_{n\nu}\nabla_\rho w_n^{\mu\nu\rho} + 3(\nabla_\alpha w_n^\mu)^2 +\label{s33_3}\\
& + \frac{3}{2}(\nabla_\mu w_n^\mu)^2 +a\big[ -\big((\mm[3]{n})^2+d-3\big)(w^{\mu\nu\rho}_n)^2 + \big(3(\mm[3]{n})^2+6(d-1)\big)(w^\mu_n)^2 \big] \Big\}\;,\notag\\
S_{32} = & 3a^2\int \sum^\infty_{n=1} \mm[3]{n-1}\Big\{ 4\nabla_\mu w_{n-1\nu} h^{\mu\nu}_{n} - 2\nabla_\mu w^{\mu\nu\rho}_{n-1}h_{n\nu\rho}  - \nabla_\mu w^\mu_{n-1}  h_{n} \Big\} \;,\\
S_{31} =& \frac{6(d+1)a^2}{d}\int\sum^\infty_{n=2} \mm{n-1}\mm[3]{n-2}w_{n-2|\mu} A^\mu_n\;,\\
S_{30} =&    -3a\int\sum^\infty_{m=0,n=1} \mm{n-1} \nabla_\mu w^\mu_m \phi_n (P^3_m,P^2_{n-1})_2\;,\notag\\
S_{22} =&  3a\int\sum^\infty_{n=0}  \Big\{ -(\nabla_\alpha h^{\mu\nu}_n)^2 + 2(\nabla_\mu h^{\mu\nu}_n)^2 -2\nabla_{\mu} h_n\nabla_\nu h_n^{\mu\nu} + (\nabla_\alpha h_n)^2 +\\
&+a\Big[ -2d(h^{\mu\nu}_n)^2 +\frac{1}{2}\big(3(\mm{n})^2+3d-2\big)(h_n)^2 \Big] \Big\}  \;,\notag\\
S_{21} =& \frac{12(d+1)a}{d}\int\sum^\infty_{n=1} \mm{n-1} \Big\{ -\nabla_\mu h^{\mu\nu}_{n-1} A_{n\nu} + \nabla_\mu h_{n-1} A^\mu_n  \Big\}\;,\\
S_{20} =&  3\int\sum^{\infty}_{m,n=0}\Big\{ 2\phi_m\big[ \nabla_\mu\nabla_\nu h_n^{\mu\nu} - \Box h_n\big](P^1_m,P^2_n)_1 +\\
&+ a\phi_m h_n\big(\big[-\cosh^2z \LL_{-2d-2}\LL_d+2(3d+1)\big]P^1_m,P^2_n)_1 \Big\} \;,\notag\\
S_{11} =& -\frac{3(d+1)}{d^2}\int\sum^\infty_{n=0} \Big\{ d\big[ (\nabla_\alpha A^\mu_n)^2 - (\nabla_\mu A^\mu_n)^2 \big] +\\
&+  a\Big[ 3d^2+d-4- (d+2)(\mm[1]{n})^2 \Big](A^\mu_n)^2 \Big\}	\;,\notag\\
S_{10} =& -\frac{6(d+1)}{d}\int \sum^\infty_{m,n=0} \nabla_\mu A^\mu_m\phi_n (P^1_m,\LL_{-d}P^1_n)_1 \;,\\
S_{00} =&  \frac{1}{a}\int \sum^\infty_{m,n=0} \Big\{ 2\nabla_\alpha\phi_m\nabla^\alpha\phi_n(P^1_m,[1-\tanh^2z]P^1_n)_1 +\label{s00_3} \\
&+\frac{a}{2} \phi_m\phi_n(P^1_m,[-\LL_{-6d}\LL_d + d(3d-1)\tanh^2z+11d+4]P^1_n)_1\Big\} \;.\notag
\end{align}
where the integration measure is introduced  
\begin{equation}
\int = {{a}^{d/2}}{{b}^{-(d+5)/2}}\int d\mu_d\;.
\end{equation}

The fields $A^\mu_n$ and $\phi_n$ with  $n=1,2,..., \infty$ can be gauged away by means of the Stueckelberg-type gauge transformations using all  vector and scalar gauge parameters $\xi^\mu_n$ and $\xi_n$ with  $n=0,1,2,..., \infty$. The modes $w^{\mu\nu\rho}_n$ and $h^{\mu\nu}_n$  remain intact. Then, the resulting partially gauged fixed action \eqref{s33_3}-\eqref{s00_3} reads
\begin{align}
S_{33} =& a^2\int\sum^\infty_{n=0}\Big\{ -(\nabla_\alpha w_n^{\mu\nu\rho})^2+3(\nabla_\mu w_n^{\mu\nu\rho})^2 - 6\nabla_\mu w_{n\nu}\nabla_\rho w_n^{\mu\nu\rho} + 3(\nabla_\alpha w_n^\mu)^2 +\label{S33_3}\\
& + \frac{3}{2}(\nabla_\mu w_n^\mu)^2 +a\big[ -\big((\mm[3]{n})^2+d-3\big)(w^{\mu\nu\rho}_n)^2 + \big(3(\mm[3]{n})^2+6(d-1)\big)(w^\mu_n)^2 \big] \Big\}\;,\notag\\
S_{32} = & 3a^2\int \sum^\infty_{n=1} \mm[3]{n-1}\Big\{ 4\nabla_\mu w_{n-1\nu} h^{\mu\nu}_{n} - 2\nabla_\mu w^{\mu\nu\rho}_{n-1}h_{n\nu\rho}  - \nabla_\mu w^\mu_{n-1}h_{n} \Big\} \;,\\
S_{31} =& 0\;, \qquad\quad S_{30} = 0 \;,\\
S_{22} =&  3a\int\sum^\infty_{n=0}  \Big\{ -(\nabla_\alpha h^{\mu\nu}_n)^2 + 2(\nabla_\mu h^{\mu\nu}_n)^2 -2\nabla_{\mu} h_n\nabla_\nu h_n^{\mu\nu} + (\nabla_\alpha h_n)^2 +\\
&+a\Big[ -2d(h^{\mu\nu}_n)^2 +\frac{1}{2}\big(3(\mm{n})^2+3d-2\big)(h_n)^2 \Big] \Big\}  \;,\notag \qquad\quad S_{21} = 0\;,\\
S_{20} =& 3\int\sum^{\infty}_{n=0}\Big\{ 2\phi_0\big[ \nabla_\mu\nabla_\nu h_n^{\mu\nu} - \Box h_n\big](P^1_0,P^2_n)_1 +\\
&+ a\phi_0 h_n\big(\big[-\cosh^2z \LL_{-2d-2}\LL_d+2(3d+1)\big]P^1_0,P^2_n\big)_1 \Big\} = \notag\\
=& 3\int\Big\{ 2\phi_0\big[ \nabla_\mu\nabla_\nu h^{\mu\nu}_0 - \Box h_0\big]\sqrt{\frac{d}{d+1}} + a\phi_0 h_0\Bigg( 0+2(3d+1)\sqrt{\frac{d}{d+1}} \Bigg) \Big\} = \notag\\
=&  6\sqrt{\frac{d}{d+1}}\int \phi_0\big[ \nabla_\mu\nabla_\nu h_0^{\mu\nu} - \Box h_0 +a(3d+1)h_0 \big] \;,\notag\\
S_{11} =& -\frac{3(d+1)}{d^2}\int \Big\{d\big[ (\nabla_\alpha A^\mu_0)^2 - (\nabla_\mu A^\mu_0)^2 \big]  + a(2d^2+d)(A^\mu_0)^2\Big\}	\;,\\
S_{10} =& -\frac{6(d+1)}{d}\int\nabla_\mu A^\mu_0\phi_0 (P^1_0,\LL_{-d}P^1_0)_1 =\\
=& -\frac{6(d+1)}{d}\int \nabla_\mu A^\mu_0\phi_0 (P^1_0,[\LL_d - 2d\tanh z]P^1_0)_1 = \\
=& -\frac{6(d+1)}{d}\int \nabla_\mu A^\mu_0\phi_0(0-2d\cdot 0)  = 0\,,\notag\\
S_{00} =&  \frac{1}{a}\int \Big\{ 2(\nabla_\alpha\phi_0)^2(P^1_0,[1-\tanh^2z]P^1_0)_1 +\label{S00_3} \\
&+\frac{a}{2} (\phi_0)^2(P^1_0,[-\LL_{-6d}\LL_d + d(3d-1)\tanh^2z+11d+4]P^1_0)_1\Big\} = \notag\\
=&\frac{1}{a}\int\Big\{ 2(\nabla_\alpha \phi_0)^2 \Big( 1-\frac{1}{d+1}\Big) +\frac{a}{2}(\phi_0)^2 \Big( 0+ d(3d-1)\cdot \frac{1}{d+1} + 11d+4 \Big) \Big\}=\notag\\
=& \frac{1}{a(d+1)}\int\Big\{ 2d(\nabla_\alpha \phi_0)^2 +a(7d^2+7d+2)(\phi_0)^2 \Big\}\;.\notag
\end{align}

Evaluating the inner products and varying  the action \eqref{S33_3}-\eqref{S00_3} one finds the equations of motion of the bottom theory:
\begin{align}
&\frac{\delta S}{\delta w^{\mu\nu\rho}_n} = 0: &&n\geq 0: && \Box w^{\mu\nu\rho}_n - \nabla^{(\mu}\nabla_\sigma w^{\nu\rho)\sigma}_n + \frac{1}{2}\nabla^{(\mu}\nabla^{\nu}w^{\rho)}_n - g^{(\mu\nu}\Box w^{\rho)}_n + \label{e1} \\
&&&&& + g^{(\mu\nu}\nabla_\sigma\nabla_\zeta w^{\rho)\sigma\zeta}_n - \frac{1}{2}g^{(\mu\nu}\nabla^{\rho)}\nabla_\sigma w^\sigma_n +\notag\\
&&&&& + a\big[-((\mm[3]{n})^2+d-3)w^{\mu\nu\rho}_n+((\mm[3]{n})^2+2(d-1))g^{(\mu\nu}w^{\rho)}_n\big] + \notag\\
&&&&&+ \mm[3]{n}\big[-2g^{(\mu\nu}\nabla_\sigma h^{\rho)\sigma}_{n+1}+\nabla^{\mu}h^{\nu\rho)}_{n+1}+\frac{1}{2}g^{(\mu\nu}\nabla^{\rho)}h_{n+1}\big] = 0\;,\notag\\
&\frac{\delta S}{\delta h^{\mu\nu}_n} = 0: && n=0: && \Box h^{\mu\nu}_0-\nabla^{(\mu}\nabla_\sigma h^{\nu)\sigma}_0+\nabla^\mu\nabla^\nu h_0 - g^{\mu\nu}\Box  h_0 + \nabla_\sigma\nabla_\rho h^{\sigma\rho}_0 + \\
&&&&& +  a\big[-2dh^{\mu\nu}_0+\frac{1}{2}(3(\mm{0})^2+3d-2)g^{\mu\nu}h_0\big] + \notag\\
&&&&& + \frac{1}{a}\sqrt{\frac{d}{d+1}}\big[\nabla^\mu\nabla^\nu - g^{\mu\nu}\Box + a(3d+1)g^{\mu\nu}\big]\phi_0 = 0\;,\\
&&& n \geq 1: && \Box h^{\mu\nu}_n-\nabla^{(\mu}\nabla_\sigma h^{\nu)\sigma}_n+\nabla^\mu\nabla^\nu h_n - g^{\mu\nu}\Box  h_n + \nabla_\sigma\nabla_\rho h^{\sigma\rho}_n + \notag\\
&&&&& + a\big[-2dh^{\mu\nu}_n+\frac{1}{2}(3(\mm{n})^2+3d-2)g^{\mu\nu}h_n\big] + \notag\\
&&&&& + a\big[ \nabla^{(\mu}w^{\nu)}_{n-1} - \nabla_\rho w^{\mu\nu\rho}_{n-1} - \frac{1}{2}g^{\mu\nu}\nabla_\rho w^{\rho}_{n-1} \big] = 0\;,\notag\\
&\frac{\delta S}{\delta A^\mu_0} = 0: &&&& \Box A^\mu_0 - \nabla^\mu\nabla_\nu A^\nu_0 - a(2d+1) A^\mu_0 = 0\;,\label{e5}\\
&\frac{\delta S}{\delta \phi_0} = 0: &&&& \left[ \Box -  \frac{a(7d^2+7d+2)}{2d}\right] \phi_0 -\\
&&&&&- \frac{3}{2}\sqrt{\frac{d+1}{d}}a\big[ \nabla_\mu\nabla_\nu - g_{\mu\nu}\Box + a(3d+1)g_{\mu\nu}\big]h^{\mu\nu}_0  = 0\;.\notag
\end{align}
Similar to the spin-2 case the vector mode  $A^\mu_0$ decouples from the other fields. Also, there is the residual gauge invariance  
\be
\label{gauge_s3}
\delta w^{\mu\nu\rho}_n = \nabla^{(\mu}\xi^{\nu\rho)}_n \,,
\qquad
\delta h^{\mu\nu}_n = a\mm[3]{n-1}\xi^{\mu\nu}_{n-1}\,,
\ee
with respect to  the traceless parameters $\xi^{\mu\nu}_n$ with $n=0,1,2,...,\infty$.  The invariance  can be used to impose the TT gauge for rank-3 fields (see Section \bref{sec:spin3}).

%\bibliographystyle{JHEP}
%\bibliography{HSmaster}

\begin{thebibliography}{10}

\bibitem{Kaluza:1921tu}
T.~Kaluza, \emph{{Zum Unit\"atsproblem der Physik}},
  \href{http://dx.doi.org/10.1142/S0218271818700017}{\emph{Sitzungsber. Preuss.
  Akad. Wiss. Berlin (Math. Phys. )} {\bf 1921} (1921) 966--972},
  [\href{http://arxiv.org/abs/1803.08616}{{\tt 1803.08616}}].

\bibitem{Klein:1926tv}
O.~Klein, \emph{{Quantum Theory and Five-Dimensional Theory of Relativity. (In
  German and English)}}, \href{http://dx.doi.org/10.1007/BF01397481}{\emph{Z.
  Phys.} {\bf 37} (1926) 895--906}.

\bibitem{Metsaev:2000qb}
R.~Metsaev, \emph{{Massive fields in AdS(3) and compactification in AdS space
  time}}, \href{http://dx.doi.org/10.1016/S0920-5632(01)01543-2}{\emph{Nucl.
  Phys. B Proc. Suppl.} {\bf 102} (2001) 100--106},
  [\href{http://arxiv.org/abs/hep-th/0103088}{{\tt hep-th/0103088}}].

\bibitem{Artsukevich:2008vy}
A.~Artsukevich and M.~Vasiliev, \emph{{On Dimensional Degression in AdS(d)}},
  \href{http://dx.doi.org/10.1103/PhysRevD.79.045007}{\emph{Phys. Rev. D} {\bf
  79} (2009) 045007}, [\href{http://arxiv.org/abs/0810.2065}{{\tt 0810.2065}}].

\bibitem{Gwak:2016sma}
S.~Gwak, J.~Kim and S.-J. Rey, \emph{{Massless and Massive Higher Spins from
  Anti-de Sitter Space Waveguide}},
  \href{http://dx.doi.org/10.1007/JHEP11(2016)024}{\emph{JHEP} {\bf 11} (2016)
  024}, [\href{http://arxiv.org/abs/1605.06526}{{\tt 1605.06526}}].

\bibitem{Brown:1986nw}
J.~D. Brown and M.~Henneaux, \emph{{Central Charges in the Canonical
  Realization of Asymptotic Symmetries: An Example from Three-Dimensional
  Gravity}}, \href{http://dx.doi.org/10.1007/BF01211590}{\emph{Commun. Math.
  Phys.} {\bf 104} (1986) 207--226}.

\bibitem{Balasubramanian:1998sn}
V.~Balasubramanian, P.~Kraus and A.~E. Lawrence, \emph{{Bulk versus boundary
  dynamics in anti-de Sitter space-time}},
  \href{http://dx.doi.org/10.1103/PhysRevD.59.046003}{\emph{Phys. Rev. D} {\bf
  59} (1999) 046003}, [\href{http://arxiv.org/abs/hep-th/9805171}{{\tt
  hep-th/9805171}}].

\bibitem{Mazac:2016qev}
D.~Mazac, \emph{{Analytic bounds and emergence of AdS$_{2}$ physics from the
  conformal bootstrap}},
  \href{http://dx.doi.org/10.1007/JHEP04(2017)146}{\emph{JHEP} {\bf 04} (2017)
  146}, [\href{http://arxiv.org/abs/1611.10060}{{\tt 1611.10060}}].

\bibitem{Anninos:2019oka}
D.~Anninos, D.~M. Hofman and J.~Kruthoff, \emph{{Charged Quantum Fields in
  AdS$_2$}},
  \href{http://dx.doi.org/10.21468/SciPostPhys.7.4.054}{\emph{SciPost Phys.}
  {\bf 7} (2019) 054}, [\href{http://arxiv.org/abs/1906.00924}{{\tt
  1906.00924}}].

\bibitem{Anous:2020nxu}
T.~Anous and J.~Skulte, \emph{{An invitation to the principal series}},
  \href{http://dx.doi.org/10.21468/SciPostPhys.9.3.028}{\emph{SciPost Phys.}
  {\bf 9} (2020) 028}, [\href{http://arxiv.org/abs/2007.04975}{{\tt
  2007.04975}}].

\bibitem{Boulanger:2014vya}
N.~Boulanger, D.~Ponomarev, E.~Sezgin and P.~Sundell, \emph{{New unfolded
  higher spin systems in $AdS_3$}},
  \href{http://dx.doi.org/10.1088/0264-9381/32/15/155002}{\emph{Class. Quant.
  Grav.} {\bf 32} (2015) 155002}, [\href{http://arxiv.org/abs/1412.8209}{{\tt
  1412.8209}}].

\bibitem{Gwak:2015jdo}
S.~Gwak, E.~Joung, K.~Mkrtchyan and S.-J. Rey, \emph{{Rainbow vacua of colored
  higher-spin (A)dS$_{3}$ gravity}},
  \href{http://dx.doi.org/10.1007/JHEP05(2016)150}{\emph{JHEP} {\bf 05} (2016)
  150}, [\href{http://arxiv.org/abs/1511.05975}{{\tt 1511.05975}}].

\bibitem{Basile:2018dzi}
T.~Basile, X.~Bekaert and E.~Joung, \emph{{Twisted Flato-Fronsdal Theorem for
  Higher-Spin Algebras}},
  \href{http://dx.doi.org/10.1007/JHEP07(2018)009}{\emph{JHEP} {\bf 07} (2018)
  009}, [\href{http://arxiv.org/abs/1802.03232}{{\tt 1802.03232}}].

\bibitem{Alkalaev:2019xuv}
K.~Alkalaev and X.~Bekaert, \emph{{Towards higher-spin AdS$_2$/CFT$_1$
  holography}}, \href{http://dx.doi.org/10.1007/JHEP04(2020)206}{\emph{JHEP}
  {\bf 04} (2020) 206}, [\href{http://arxiv.org/abs/1911.13212}{{\tt
  1911.13212}}].

\bibitem{Bak:2003jk}
D.~Bak, M.~Gutperle and S.~Hirano, \emph{{A Dilatonic deformation of AdS(5) and
  its field theory dual}},
  \href{http://dx.doi.org/10.1088/1126-6708/2003/05/072}{\emph{JHEP} {\bf 05}
  (2003) 072}, [\href{http://arxiv.org/abs/hep-th/0304129}{{\tt
  hep-th/0304129}}].

\bibitem{Hinterbichler:2013kwa}
K.~Hinterbichler, J.~Levin and C.~Zukowski, \emph{{Kaluza-Klein Towers on
  General Manifolds}},
  \href{http://dx.doi.org/10.1103/PhysRevD.89.086007}{\emph{Phys. Rev. D} {\bf
  89} (2014) 086007}, [\href{http://arxiv.org/abs/1310.6353}{{\tt 1310.6353}}].

\bibitem{Gross:2017aos}
D.~J. Gross and V.~Rosenhaus, \emph{{All point correlation functions in SYK}},
  \href{http://dx.doi.org/10.1007/JHEP12(2017)148}{\emph{JHEP} {\bf 12} (2017)
  148}, [\href{http://arxiv.org/abs/1710.08113}{{\tt 1710.08113}}].

\bibitem{Gutperle:2020gez}
M.~Gutperle and C.~F. Uhlemann, \emph{{Janus on the Brane}},
  \href{http://dx.doi.org/10.1007/JHEP07(2020)243}{\emph{JHEP} {\bf 07} (2020)
  243}, [\href{http://arxiv.org/abs/2003.12080}{{\tt 2003.12080}}].

\bibitem{Fronsdal:1978rb}
C.~Fronsdal, \emph{{Massless Fields with Integer Spin}},
  \href{http://dx.doi.org/10.1103/PhysRevD.18.3624}{\emph{Phys. Rev.} {\bf D18}
  (1978) 3624}.

\bibitem{Lopatin:1988hz}
V.~E. Lopatin and M.~A. Vasiliev, \emph{Free massless bosonic fields of
  arbitrary spin in d- dimensional de {S}itter space}, {\emph{Mod. Phys. Lett.}
  {\bf A3} (1988) 257}.

\bibitem{Buchbinder:2001bs}
I.~L. Buchbinder, A.~Pashnev and M.~Tsulaia, \emph{{Lagrangian formulation of
  the massless higher integer spin fields in the AdS background}},
  \href{http://dx.doi.org/10.1016/S0370-2693(01)01268-0}{\emph{Phys. Lett. B}
  {\bf 523} (2001) 338--346}, [\href{http://arxiv.org/abs/hep-th/0109067}{{\tt
  hep-th/0109067}}].

\bibitem{Blencowe:1988gj}
M.~P. Blencowe, \emph{{A Consistent Interacting Massless Higher Spin Field
  Theory in $D$ = (2+1)}},
  \href{http://dx.doi.org/10.1088/0264-9381/6/4/005}{\emph{Class. Quant. Grav.}
  {\bf 6} (1989) 443}.

\bibitem{Campoleoni:2010zq}
A.~Campoleoni, S.~Fredenhagen, S.~Pfenninger and S.~Theisen, \emph{{Asymptotic
  symmetries of three-dimensional gravity coupled to higher-spin fields}},
  \href{http://dx.doi.org/10.1007/JHEP11(2010)007}{\emph{JHEP} {\bf 11} (2010)
  007}, [\href{http://arxiv.org/abs/1008.4744}{{\tt 1008.4744}}].

\bibitem{Campoleoni:2011tn}
A.~Campoleoni, \emph{{Higher Spins in D = 2 + 1}},
  \href{http://dx.doi.org/10.1142/9789814522519_0020}{\emph{Subnucl. Ser.} {\bf
  49} (2013) 385--396}, [\href{http://arxiv.org/abs/1110.5841}{{\tt
  1110.5841}}].

\bibitem{Hinterbichler:2011tt}
K.~Hinterbichler, \emph{{Theoretical Aspects of Massive Gravity}},
  \href{http://dx.doi.org/10.1103/RevModPhys.84.671}{\emph{Rev. Mod. Phys.}
  {\bf 84} (2012) 671--710}, [\href{http://arxiv.org/abs/1105.3735}{{\tt
  1105.3735}}].

\bibitem{Metsaev:1997nj}
R.~R. Metsaev, \emph{{Arbitrary spin massless bosonic fields in d-dimensional
  anti-de Sitter space}},  \href{http://arxiv.org/abs/hep-th/9810231}{{\tt
  hep-th/9810231}}.

\bibitem{Nicolai:1984hb}
H.~Nicolai, \emph{{Representations of supersymmetry in anti-de sitter space}},
  in \emph{{Spring School on Supergravity and Supersymmetry}}, 4, 1984.

\bibitem{deWit:1999ui}
B.~de~Wit and I.~Herger, \emph{{Anti-de Sitter supersymmetry}}, {\emph{Lect.
  Notes Phys.} {\bf 541} (2000) 79--100},
  [\href{http://arxiv.org/abs/hep-th/9908005}{{\tt hep-th/9908005}}].

\bibitem{Deser:2001pe}
S.~Deser and A.~Waldron, \emph{Gauge invariances and phases of massive higher
  spins in {(A)dS}}, {\emph{Phys. Rev. Lett.} {\bf 87} (2001) 031601},
  [\href{http://arxiv.org/abs/hep-th/0102166}{{\tt hep-th/0102166}}].

\bibitem{Bergshoeff:2009hq}
E.~A. Bergshoeff, O.~Hohm and P.~K. Townsend, \emph{{Massive Gravity in Three
  Dimensions}},
  \href{http://dx.doi.org/10.1103/PhysRevLett.102.201301}{\emph{Phys. Rev.
  Lett.} {\bf 102} (2009) 201301}, [\href{http://arxiv.org/abs/0901.1766}{{\tt
  0901.1766}}].

\bibitem{Gelfand1966}
I.~Gelfand and A.~Kirillov, \emph{{Sur les corps liés aux algèbres
  enveloppantes des algèbres de Lie}}, {\emph{Publications Mathématiques de
  l'IHÉS} {\bf 31} (1966) 5--19}.

\bibitem{Joung:2014qya}
E.~Joung and K.~Mkrtchyan, \emph{{Notes on higher-spin algebras: minimal
  representations and structure constants}},
  \href{http://dx.doi.org/10.1007/JHEP05(2014)103}{\emph{JHEP} {\bf 05} (2014)
  103}, [\href{http://arxiv.org/abs/1401.7977}{{\tt 1401.7977}}].

\bibitem{Joung:2019wwf}
E.~Joung, K.~Mkrtchyan and G.~Poghosyan, \emph{{Looking for partially-massless
  gravity}}, \href{http://dx.doi.org/10.1007/JHEP07(2019)116}{\emph{JHEP} {\bf
  07} (2019) 116}, [\href{http://arxiv.org/abs/1904.05915}{{\tt 1904.05915}}].

\bibitem{Achucarro:1993fd}
A.~Achucarro and M.~E. Ortiz, \emph{{Relating black holes in two-dimensions and
  three-dimensions}},
  \href{http://dx.doi.org/10.1103/PhysRevD.48.3600}{\emph{Phys. Rev. D} {\bf
  48} (1993) 3600--3605}, [\href{http://arxiv.org/abs/hep-th/9304068}{{\tt
  hep-th/9304068}}].

\bibitem{Guralnik:2003we}
G.~Guralnik, A.~Iorio, R.~Jackiw and S.~Pi, \emph{{Dimensionally reduced
  gravitational Chern-Simons term and its kink}},
  \href{http://dx.doi.org/10.1016/S0003-4916(03)00142-8}{\emph{Annals Phys.}
  {\bf 308} (2003) 222--236}, [\href{http://arxiv.org/abs/hep-th/0305117}{{\tt
  hep-th/0305117}}].

\bibitem{Grumiller:2003ad}
D.~Grumiller and W.~Kummer, \emph{{The Classical solutions of the dimensionally
  reduced gravitational Chern-Simons theory}},
  \href{http://dx.doi.org/10.1016/S0003-4916(03)00138-6}{\emph{Annals Phys.}
  {\bf 308} (2003) 211--221}, [\href{http://arxiv.org/abs/hep-th/0306036}{{\tt
  hep-th/0306036}}].

\bibitem{Mertens:2018fds}
T.~G. Mertens, \emph{{The Schwarzian theory \textemdash{} origins}},
  \href{http://dx.doi.org/10.1007/JHEP05(2018)036}{\emph{JHEP} {\bf 05} (2018)
  036}, [\href{http://arxiv.org/abs/1801.09605}{{\tt 1801.09605}}].

\bibitem{Gaikwad:2018dfc}
A.~Gaikwad, L.~K. Joshi, G.~Mandal and S.~R. Wadia, \emph{{Holographic dual to
  charged SYK from 3D Gravity and Chern-Simons}},
  \href{http://dx.doi.org/10.1007/JHEP02(2020)033}{\emph{JHEP} {\bf 02} (2020)
  033}, [\href{http://arxiv.org/abs/1802.07746}{{\tt 1802.07746}}].

\bibitem{Narayan:2020pyj}
K.~Narayan, \emph{{On aspects of 2-dim dilaton gravity, dimensional reduction
  and holography}},  \href{http://arxiv.org/abs/2010.12955}{{\tt 2010.12955}}.

\bibitem{Henneaux:2010xg}
M.~Henneaux and S.-J. Rey, \emph{{Nonlinear W(infinity) Algebra as Asymptotic
  Symmetry of Three-Dimensional Higher Spin Anti-de Sitter Gravity}},
  \href{http://arxiv.org/abs/1008.4579}{{\tt 1008.4579}}.

\bibitem{Campoleoni:2012hp}
A.~Campoleoni, S.~Fredenhagen, S.~Pfenninger and S.~Theisen, \emph{{Towards
  metric-like higher-spin gauge theories in three dimensions}},
  \href{http://dx.doi.org/10.1088/1751-8113/46/21/214017}{\emph{J. Phys.} {\bf
  A46} (2013) 214017}, [\href{http://arxiv.org/abs/1208.1851}{{\tt
  1208.1851}}].

\bibitem{Fredenhagen:2014oua}
S.~Fredenhagen and P.~Kessel, \emph{{Metric- and frame-like higher-spin gauge
  theories in three dimensions}},
  \href{http://dx.doi.org/10.1088/1751-8113/48/3/035402}{\emph{J. Phys.} {\bf
  A48} (2015) 035402}, [\href{http://arxiv.org/abs/1408.2712}{{\tt
  1408.2712}}].

\bibitem{Alkalaev:2013fsa}
K.~B. Alkalaev, \emph{{On higher spin extension of the Jackiw-Teitelboim
  gravity model}},
  \href{http://dx.doi.org/10.1088/1751-8113/47/36/365401}{\emph{J. Phys.} {\bf
  A47} (2014) 365401}, [\href{http://arxiv.org/abs/1311.5119}{{\tt
  1311.5119}}].

\bibitem{Grumiller:2013swa}
D.~Grumiller, M.~Leston and D.~Vassilevich, \emph{{Anti-de Sitter holography
  for gravity and higher spin theories in two dimensions}},
  \href{http://dx.doi.org/10.1103/PhysRevD.89.044001}{\emph{Phys. Rev. D} {\bf
  89} (2014) 044001}, [\href{http://arxiv.org/abs/1311.7413}{{\tt 1311.7413}}].

\bibitem{Alkalaev:2014qpa}
K.~B. Alkalaev, \emph{{Global and local properties of AdS$_{2}$ higher spin
  gravity}}, \href{http://dx.doi.org/10.1007/JHEP10(2014)122}{\emph{JHEP} {\bf
  10} (2014) 122}, [\href{http://arxiv.org/abs/1404.5330}{{\tt 1404.5330}}].

\bibitem{Datta:2021efl}
S.~Datta, \emph{{The Schwarzian sector of higher spin CFTs}},
  \href{http://dx.doi.org/10.1007/JHEP04(2021)171}{\emph{JHEP} {\bf 04} (2021)
  171}, [\href{http://arxiv.org/abs/2101.04980}{{\tt 2101.04980}}].

\bibitem{Alkalaev:2020kut}
K.~Alkalaev and X.~Bekaert, \emph{{On BF-type higher-spin actions in two
  dimensions}}, \href{http://dx.doi.org/10.1007/JHEP05(2020)158}{\emph{JHEP}
  {\bf 05} (2020) 158}, [\href{http://arxiv.org/abs/2002.02387}{{\tt
  2002.02387}}].

\bibitem{Gross:2017hcz}
D.~J. Gross and V.~Rosenhaus, \emph{{The Bulk Dual of SYK: Cubic Couplings}},
  \href{http://dx.doi.org/10.1007/JHEP05(2017)092}{\emph{JHEP} {\bf 05} (2017)
  092}, [\href{http://arxiv.org/abs/1702.08016}{{\tt 1702.08016}}].

\bibitem{Bonifacio:2019ioc}
J.~Bonifacio and K.~Hinterbichler, \emph{{Unitarization from Geometry}},
  \href{http://dx.doi.org/10.1007/JHEP12(2019)165}{\emph{JHEP} {\bf 12} (2019)
  165}, [\href{http://arxiv.org/abs/1910.04767}{{\tt 1910.04767}}].

\bibitem{Bonifacio:2020xoc}
J.~Bonifacio and K.~Hinterbichler, \emph{{Bootstrap Bounds on Closed Einstein
  Manifolds}}, \href{http://dx.doi.org/10.1007/JHEP10(2020)069}{\emph{JHEP}
  {\bf 10} (2020) 069}, [\href{http://arxiv.org/abs/2007.10337}{{\tt
  2007.10337}}].

\bibitem{BUCHBINDER2012243}
I.~Buchbinder, T.~Snegirev and Y.~Zinoviev, \emph{Gauge invariant lagrangian
  formulation of massive higher spin fields in (a)ds3 space},
  \href{http://dx.doi.org/https://doi.org/10.1016/j.physletb.2012.08.022}{\emph{Physics
  Letters B} {\bf 716} (2012) 243--248}.

\bibitem{PhysRevD.102.066003}
M.~Grigoriev, K.~Mkrtchyan and E.~Skvortsov, \emph{Matter-free higher spin
  gravities in 3d: Partially-massless fields and general structure},
  \href{http://dx.doi.org/10.1103/PhysRevD.102.066003}{\emph{Phys. Rev. D} {\bf
  102} (Sep, 2020) 066003}.

\bibitem{Zinoviev:2017rnj}
{\relax Yu}.~M. Zinoviev, \emph{{Infinite spin fields in d = 3 and beyond}},
  \href{http://dx.doi.org/10.3390/universe3030063}{\emph{Universe} {\bf 3}
  (2017) 63}, [\href{http://arxiv.org/abs/1707.08832}{{\tt 1707.08832}}].

\bibitem{wigner1931gruppentheorie}
E.~Wigner, \emph{{Gruppentheorie und ihre Anwendung auf die Quantenmechanik der
  Atomspektren}}.
\newblock Die Wissenschaft : Sammlung von Einzelderstellungen aus den Gebieten
  der Naturwissenschaft und der Technik. F. Vieweg \& Sohn Akt.-Ges., 1931.

\bibitem{Bargmann:1946me}
V.~Bargmann, \emph{{Irreducible unitary representations of the Lorentz group}},
  \href{http://dx.doi.org/10.2307/1969129}{\emph{Annals Math.} {\bf 48} (1947)
  568--640}.

\bibitem{Barut1965}
A.~O. Barut and C.~Fronsdal, \emph{{On non-compact groups. II. Representations
  of the 2+1 Lorentz group}},
  \href{http://dx.doi.org/10.1098/rspa.1965.0195}{\emph{Proceedings of the
  Royal Society of London. Series A. Mathematical and Physical Sciences} {\bf
  287} (1965) 532--548}.

\bibitem{Pukanszky1961}
L.~Pukánszky, \emph{{On the Kronecker Products of Irreducible Representations
  of the $2\times 2$ Real Unimodular Group. I}}, {\emph{Transactions of the
  American Mathematical Society} {\bf 100} (1961) 116--152}.

\bibitem{Mukunda:1974gb}
N.~Mukunda and B.~Radhakrishnan, \emph{{Clebsch-Gordan problem and coefficients
  for the three-dimensional Lorentz group in a continuous basis. 1}},
  \href{http://dx.doi.org/10.1063/1.1666814}{\emph{J. Math. Phys.} {\bf 15}
  (1974) 1320--1331}.

\bibitem{repka}
J.~Repka, \emph{{Tensor Products of Unitary Representations of $sl(2,R)$}},
  {\emph{American Journal of Mathematics} {\bf 100} (1978) 747--774}.

\bibitem{zbMATH03701150}
V.~F. {Molchanov}, \emph{{Tensor products of unitary representations of the
  three-dimensional Lorentz group}},
  \href{http://dx.doi.org/10.1070/IM1980v015n01ABEH001191}{\emph{{Math. USSR,
  Izv.}} {\bf 15} (1980) 113--143}.

\bibitem{vanTonder:2002gh}
A.~van Tonder, \emph{{Cohomology and decomposition of tensor product
  representations of SL(2,R)}},
  \href{http://dx.doi.org/10.1016/j.nuclphysb.2003.10.029}{\emph{Nucl. Phys. B}
  {\bf 677} (2004) 614--632}, [\href{http://arxiv.org/abs/hep-th/0212149}{{\tt
  hep-th/0212149}}].

\bibitem{Sellaroli:2014ega}
G.~Sellaroli, \emph{{Wigner-Eckart theorem for the non-compact algebra
  $sl(2,R)$}}, \href{http://dx.doi.org/10.1063/1.4916889}{\emph{J. Math. Phys.}
  {\bf 56} (2015) 041701}, [\href{http://arxiv.org/abs/1411.7467}{{\tt
  1411.7467}}].

\bibitem{Shen2011}
J.~Shen, T.~Tang and L.-L. Wang, \emph{Orthogonal Polynomials and Related
  Approximation Results}, pp.~47--140.
\newblock Springer Berlin Heidelberg, Berlin, Heidelberg, 2011.

\bibitem{Conde:2016izb}
E.~Conde, E.~Joung and K.~Mkrtchyan, \emph{{Spinor-Helicity Three-Point
  Amplitudes from Local Cubic Interactions}},
  \href{http://dx.doi.org/10.1007/JHEP08(2016)040}{\emph{JHEP} {\bf 08} (2016)
  040}, [\href{http://arxiv.org/abs/1605.07402}{{\tt 1605.07402}}].

\bibitem{Mkrtchyan:2017ixk}
K.~Mkrtchyan, \emph{{Cubic interactions of massless bosonic fields in three
  dimensions}},
  \href{http://dx.doi.org/10.1103/PhysRevLett.120.221601}{\emph{Phys. Rev.
  Lett.} {\bf 120} (2018) 221601}, [\href{http://arxiv.org/abs/1712.10003}{{\tt
  1712.10003}}].

\bibitem{Kessel:2018ugi}
P.~Kessel and K.~Mkrtchyan, \emph{{Cubic interactions of massless bosonic
  fields in three dimensions II: Parity-odd and Chern-Simons vertices}},
  \href{http://dx.doi.org/10.1103/PhysRevD.97.106021}{\emph{Phys. Rev. D} {\bf
  97} (2018) 106021}, [\href{http://arxiv.org/abs/1803.02737}{{\tt
  1803.02737}}].

\end{thebibliography}

\providecommand{\href}[2]{#2}\begingroup\raggedright\endgroup

\end{document}